\documentclass[11pt,draftcls,onecolumn,twoside]{IEEEtran}

\usepackage{amsmath}
\usepackage{amssymb}
\usepackage{latexsym}
\usepackage{verbatim}
\usepackage[final]{graphicx}
\usepackage{psfrag}

\newtheorem{proposition}{Proposition}
\newtheorem{lemma}{Lemma}
\newtheorem{corollary}{Corollary}
\newtheorem{remark}{Remark}

\newtheorem{definition}{Definition}
\newtheorem{example}{Example}

{\begin{list}               
    {$\bullet$ \hfill}{
        \setlength{\leftmargin}{\parindent}
        \setlength{\parsep}{0.04\baselineskip}
        \setlength{\itemsep}{0.5\parsep}
        \setlength{\labelwidth}{\leftmargin}
        \setlength{\labelsep}{0em}}
    }
{\end{list}}

\providecommand{\eref}[1]{\eqref{eq:#1}}  
\providecommand{\cref}[1]{Chapter~\ref{chap:#1}}
\providecommand{\sref}[1]{Section~\ref{sec:#1}}
\providecommand{\fref}[1]{Figure~\ref{fig:#1}}

\providecommand{\R}{\ensuremath{\mathbb{R}}}
\providecommand{\C}{\ensuremath{\mathbb{C}}}

\providecommand{\Z}{\ensuremath{\mathbb{Z}}}

\providecommand{\abs}[1]{\lvert#1\rvert}
\providecommand{\norm}[1]{\lVert#1\rVert}

\providecommand{\set}[1]{\left\{#1\right\}}

\providecommand{\bydef}{\overset{\text{def}}{=}}

\renewcommand{\vec}[1]{\ensuremath{\boldsymbol{#1}}}
\providecommand{\mat}[1]{\ensuremath{\boldsymbol{#1}}}

\providecommand{\mA}{\mat{A}} \providecommand{\mB}{\mat{B}}
 \providecommand{\mD}{\mat{D}}
\providecommand{\mF}{\mat{F}}\providecommand{\mH}{\mat{H}}
\providecommand{\mI}{\mat{I}}

\providecommand{\mS}{\mat{S}}  
\providecommand{\mV}{\mat{V}} \providecommand{\mT}{\mat{T}}

\providecommand{\vc}{\vec{c}} \providecommand{\vd}{\vec{d}} 
\providecommand{\ve}{\vec{e}}
\providecommand{\vh}{\vec{h}} 
  
 \providecommand{\vp}{\vec{p}}
 
\providecommand{\vr}{\vec{r}} \providecommand{\vs}{\vec{s}}

\providecommand{\vx}{\vec{x}} \providecommand{\vy}{\vec{y}}
\providecommand{\vz}{\vec{z}} 
 \providecommand{\vzero}{\vec{0}}
 \providecommand{\vv}{\vec{v}}

\usepackage[nosort]{cite}

\usepackage[final]{epsfig}
\usepackage{subfigure}

\usepackage{algorithm, algorithmic}

\def\l{{\ell}}

\begin{document}

\title{On Sparse Representation in Fourier and\\ Local Bases}

\author{%
Pier Luigi Dragotti,~\IEEEmembership{Senior Member,~IEEE}~and~Yue M. Lu,~\IEEEmembership{Senior Member,~IEEE}
\thanks{P. L. Dragotti is with the Department of Electrical and Electronic Engineering, Imperial College London, London SW7 2AZ, UK (e-mail: p.dragotti@imperial.ac.uk).}%
\thanks{Y. M. Lu is with the Signals, Information, and Networks Group (SING) at the School of Engineering and Applied Sciences, Harvard University, Cambridge, MA 02138, USA (e-mail: yuelu@seas.harvard.edu).}
\thanks{This paper was presented in part at the 2013 Signal Processing with Adaptive Sparse Structured Representations (SPARS) Workshop.}
}

\markboth{}{Dragotti and Lu: Sparse Representations in Fourier and Local Bases}

\maketitle

\begin{abstract}
We consider the classical problem of finding the sparse representation of a signal in  a pair of  bases.
When both bases are orthogonal, it is known that the sparse representation is unique when the sparsity $K$ of the signal satisfies $K<1/\mu(\mD)$, where $\mu(\mD)$ is the mutual coherence of the dictionary. Furthermore, the sparse representation can be obtained in polynomial time by Basis Pursuit (BP), when $K<0.91/\mu(\mD)$. Therefore, there is a gap between the unicity condition and the one required to use the polynomial-complexity BP formulation. For the case of general dictionaries,  it is also well known that finding the sparse representation under the only constraint of unicity is NP-hard.

In this paper, we introduce, for the case of Fourier and canonical bases, a polynomial complexity algorithm that finds all the possible $K$-sparse representations of a signal under the weaker condition that $K<\sqrt{2} /\mu(\mD)$. Consequently, when $K<1/\mu(\mD)$,  the proposed algorithm solves the unique sparse representation problem for this structured dictionary in polynomial time. We further show that the same method can be extended to many other pairs of bases, one of which must have local atoms. Examples include the union of Fourier and local Fourier bases, the union of discrete cosine transform and canonical bases, and the union of random Gaussian and canonical bases.

\end{abstract}


%
%
\section{Introduction}
Consider the problem of finding the sparse representation of a signal in the union of  two orthogonal bases.
Specifically, let $\vy$ be an
 $N$-dimensional vector 
given by the linear combination
of $K$ atoms of the dictionary 
$\mD= [\mat{\Psi}, \mat{\Phi}]$, where $\mat{\Psi}$ and $\mat{\Phi}$ are two $N \times N$ orthogonal matrices.
Given the synthesis model
\begin{equation}
\vy=\mD\vx, \label{eq:SparseRepresentation}
\end{equation}
we study the problem of finding the $K$ nonzero entries of
$\vx$ from $\vy$.

One way to retrieve the sparse vector $\vx$ is to solve the following problem:
$$
(P_0): \hspace{5mm} \text{arg}\,\min_{\widetilde{\vx}} \, \norm{\widetilde{\vx}}_0 \hspace{5mm}
\mbox{s.t.} \hspace{5mm} \vy=\mD \widetilde{\vx},
$$
where the $\ell_0$
``norm'' is given by $\norm{\widetilde{\vx}}_0=\# \{i: \abs{\widetilde{x}_i} \neq 0\}$. The $(P_0)$ problem is clearly daunting since the $\ell_0$ norm is nonconvex. Therefore it might be convenient to consider the following
convex relaxation:
$$
(P_1): \hspace{5mm} \text{arg}\,\min_{\widetilde{\vx}} \| \widetilde{\vx} \|_1 \hspace{5mm}
\mbox{s.t.} \hspace{5mm} \vy=\mD \widetilde{\vx},
$$
where $\| \widetilde{\vx} \|_1= \sum_{i=1}^M \abs{\widetilde{x}_i}$ is the $\ell_1$ norm. We note that $(P_1)$, also known as Basis-Pursuit (BP)~\cite{ChenDS:98}, can be solved using polynomial complexity algorithms. 


The sparse representation problem was first posed in the above forms by
Donoho and Huo in~\cite{DonohoH:01} for the union of Fourier and canonical bases.
Specifically, let $\mu(\mD)$ denote the mutual coherence of $\mD$, defined as
$$
\mu(\mD)= \max_{1 \leq k,\ell \leq 2N, k \neq \ell} \frac{\abs{\vd_k^\ast
\vd_\ell}}{\norm{\vd_k}_2\, \norm{\vd_\ell}_2},
$$
where $\vd_k$ is the $k$th column of
$\mD$ and $(\cdot)^\ast$ denotes the conjugate transpose of a vector.
 They first showed that the original $K$-sparse vector $\vx$ is the unique solution of $(P_0)$ when
\begin{equation}
K <\frac{1}{\mu(\mD)}=\sqrt{N}, \label{eq:unique_solution}
\end{equation}
where we have used the fact that for the case of Fourier and canonical bases  $\mu(\mD)=1/\sqrt{N}$. 
They then went on showing that $(P_0)$ and $(P_1)$ are
equivalent when
\begin{equation}
K < \frac{\sqrt{N}}{2}. \label{eq:l1_constraint}
\end{equation}
This fact has important implications since it indicates that under
the constraint (\ref{eq:l1_constraint}), the sparse representation
problem has a unique solution and, more importantly, it can be solved exactly using algorithms with polynomial
complexity.

The findings of Donoho and Huo were extended to generic orthogonal pairs of bases 
by Elad and Bruckstein in~\cite{EladB:02}, where the bound in~(\ref{eq:l1_constraint}) was also improved.
Specifically, $(P_0)$  has a unique
solution, which is also equal to $\vx$, when
\begin{equation}
K < \frac{1}{\mu(\mD)}.  \label{eq:P_0_mutual}
\end{equation}
Moreover, 
if the signal $\vy$ is made of $K_p$ atoms of $\mat{\Psi}$ and $K_q$ atoms of $\mat{\Phi}$,  with $K=K_p+K_q$, then it was shown in~\cite{EladB:02} that
$(P_1)$ is equivalent to $(P_0)$
when 
\begin{equation}
2\mu(\mD)^2K_pK_q+\mu(\mD)\max\set{K_p, K_q}-1<0.
\label{eq:tight_bound}
\end{equation}
This bound is tight as demonstrated in~\cite{FeuerN:03} (see, also, Appendix~\ref{appendix:counter}), but it is
a bit obscure. For this reason, a simpler but slightly more restrictive version is 
usually adopted:
\begin{equation}
K =K_p+K_q< \frac{\sqrt{2} -0.5}{\mu(\mD)}.
\label{eq:P_0_P_1_mutual}
\end{equation}

\fref{VariousBounds:1} presents a comparison between  the $(P_0)$ bound \eref{P_0_mutual}, the tight $(P_1)$ bound \eref{tight_bound} and 
its simplified version \eref{P_0_P_1_mutual}. We immediately note that \eref{P_0_mutual} poses a weaker condition than \eref{tight_bound} or \eref{P_0_P_1_mutual}, as there is still a (small) gap between the $(P_0)$ and  $(P_1)$ bounds. While we know that $(P_1)$ can be solved with polynomial complexity algorithms, we cannot conclude from existing results whether $(P_0)$ has the same complexity, unless the sparsity level is further reduced to satisfy \eref{tight_bound}. For arbitrary redundant dictionaries,
it is well known that $(P_0)$ is NP-hard~\cite{DavisMA:97,Natarajan:95}. However, this general result does not address the case of structured dictionaries which we will be considering in this work. Moreover, another open issue is whether we can still reconstruct the vector $\vx$ when its sparsity level $K$ is beyond the $(P_0)$ bound \eref{P_0_mutual}.
\begin{figure}[t]
	\centering
	\subfigure[]{\label{fig:VariousBounds:1}
		\centering
		\includegraphics[width=0.4\textwidth]{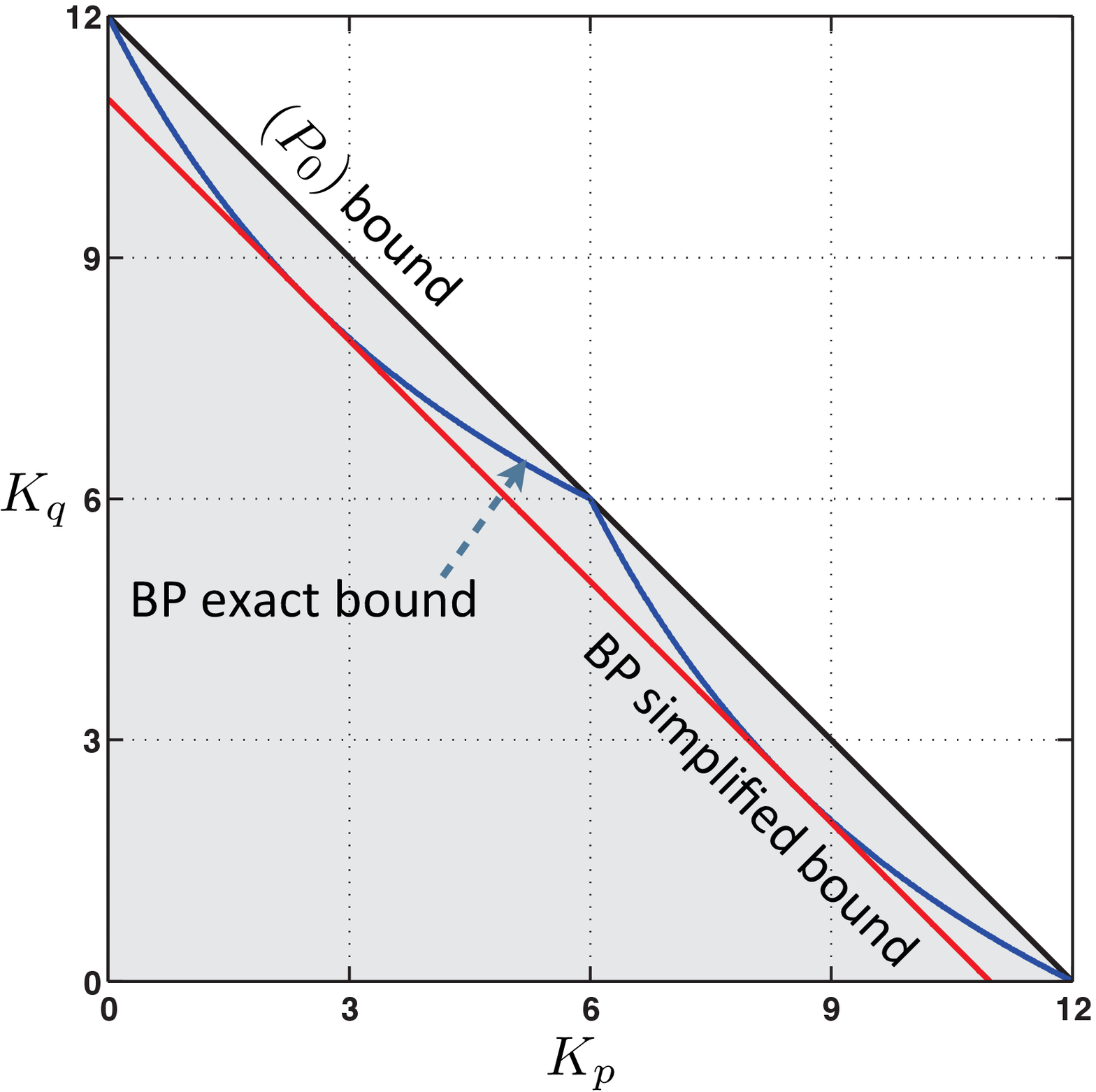}
		\vspace{-2ex}
	}
	\hspace{6ex}
	\subfigure[]{\label{fig:VariousBounds:2}
		\centering
		\includegraphics[width=0.4\textwidth]{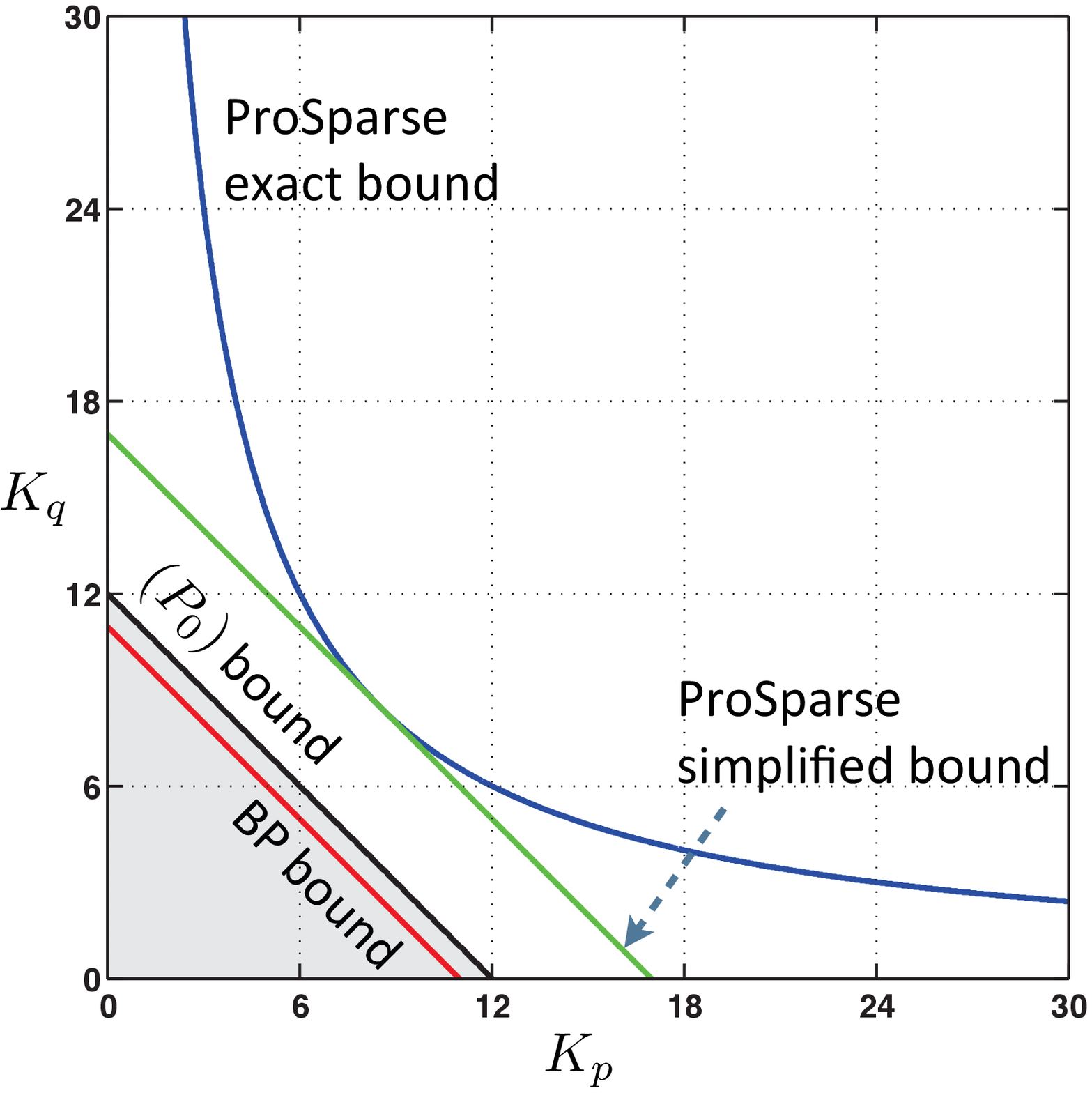}
		\vspace{-2ex}
	}
	\vspace{-2ex}
\caption{Comparing different bounds for sparse signal representation in a union of orthogonal bases. (a) Uniqueness of $(P_0)$ and the two $\ell_1$ bounds for a dictionary with $\mu(\mD)=1/12$. (b) The two ProSparse bounds \eref{prosparse_bound} and \eref{prosparse_simple_bound} plotted against the $(P_0)$ bound and the BP simplified bound, for the specific case when $\mD$ is the union of Fourier and canonical bases. We set $N = 144$, so $\mu(\mD) = 1/12$.}
\end{figure}


The main contribution of this paper is to 
show that, when $\mD$ is the union of Fourier and canonical bases, there exists a polynomial
complexity algorithm that can recover $\vx$ from $\vy = \mD \vx$, provided that
\begin{equation}\label{eq:prosparse_bound}
K_p K_q < N / 2.
\end{equation}
The proposed algorithm is based around 
Prony's method which is commonly used in spectral estimation theory~\cite{StoicaM:05}.
For this reason we name it
{\em ProSparse}---Prony's based sparsity--- in honour of Baron de Prony who first invented the method that goes under his name.

Using the inequality $2\sqrt{K_p K_q} \le K_p + K_q$, we see that a more restrictive version of \eref{prosparse_bound} is to require
\begin{equation}\label{eq:prosparse_simple_bound}
K = K_p + K_q < \sqrt{2N},
\end{equation}
which imposes a simple constraint on the total sparsity of $\vx$. In \fref{VariousBounds:2}, we compare the ProSparse bounds \eref{prosparse_bound} and \eref{prosparse_simple_bound} against the $(P_0)$ and $(P_1)$ bounds. To compute the latter two, we use the fact that $\mu(\mD)=1/\sqrt{N}$ for the case of Fourier and canonical bases. Consequently, the $(P_0)$ problem has a unique solution when the constraint \eref{unique_solution} is met and $(P_1)$ and $(P_0)$ are equivalent when
\begin{equation}
K < (\sqrt{2}-0.5)\sqrt{N}. \label{eq:l1_constraint_final}
\end{equation}
We see from the figure that the ProSparse bounds are much weaker, meaning that the proposed algorithm can recover a larger class of sparse signals. In particular, since the uniqueness bound for $(P_0)$ falls entirely within the ProSparse bounds, our results imply that, for the union of Fourier and canonical bases, the nonconvex problem $(P_0)$ can be solved with polynomial complexity under the uniqueness bound $K < \sqrt{N}$. To our knowledge, no other polynomial complexity algorithm has been known in the literature to achieve this task.

We conclude by noting that recently a generalized version of the uncertainty principle of Elad-Bruckstein was presented in~\cite{RicaudT:13} leading to more general uniqueness bounds. Those bounds converge to  \eref{unique_solution} for the case of Fourier and canonical bases. We also note that,
while finding the sparse representation of a signal is an interesting theoretical problem, modeling signals as sparse in a certain domain has proved very useful also in many signal processing applications and we refer to the paper~\cite{RicaudT:13b} and the book~\cite{Elad:10} for a comprehensive review of both theoretical as well as applied aspects of this topic. 

The rest of the paper is organized as follows: After a brief overview of Prony's method in \sref{Prony}, we present in \sref{ProSparse} the main results of this work: There we introduce ProSparse and show that it solves the sparse representation problem under the bound given in \eref{prosparse_bound}, when $\mD$ is the union of the Fourier and canonical bases. These results are then generalized in \sref{Extensions}, where we show that ProSparse works for many other pairs of bases. In general, it is only required that one of the bases have local atoms and the other basis be such that it allows for the efficient reconstruction of sparse signals from any small blocks of consecutive elements in the transform domain (see Proposition~\ref{prop:other_bases} for details). Examples of such pairs include the union of Fourier and local Fourier bases,  the union of discrete cosine transform (DCT) and canonical bases, and the union of random Gaussian and canonical bases. We conclude in \sref{Conclusions}. Unless stated otherwise, we assume throughout the paper that the basis matrices and the signals are all complex-valued, \emph{i.e.}, $\mat{\Psi}, \mat{\Phi} \in \C^{N \times N}$, $\vy \in \C^N$, and $\vx \in \C^{2N}$.

%
%
\section{Overview of Prony's method}
\label{sec:Prony} 

Consider the case when the signal $\vy$ is made only of $K$ Fourier atoms, \emph{i.e.}, $\vy = \mF \vc$, where $\mF$ is the $N$-point DFT matrix and $\vc$ is some $K$-sparse vector in $\C^N$. The algebraic structure of the Fourier matrix makes it possible to reconstruct the sparse vector $\vc$ from only $2K$ \emph{consecutive} entries of $\vy$.

One classical algorithm for such reconstruction is a method by Baron de Prony, developed in 1795 for the original purpose of estimating the  frequency, phase, and amplitude parameters of a finite sum of sinusoids \cite{Prony:1795}. In the last several decades, Prony's method has been rediscovered and extended many times in different fields: it has been used in error correcting codes (\emph{e.g.}, Berlekamp-Massey algorithm~\cite{Berlekamp:68,Massey:69}), in array signal processing~\cite{StoicaM:05}, to solve some inverse problems~\cite{MilanfarVKW:95,GustafssonHMP:00,EladMG:04}, and more recently, in parametric sampling theory~\cite{DragottiVB:07,VetterliMB:02}. 

In what follows, we present a simple derivation of the basic Prony's method, with emphasis on key results that will be used in later sections. We refer readers to the book~\cite{StoicaM:05} and to the insightful overview~\cite{EladMG:04} for more details on this intriguing nonlinear estimation algorithm and its various extensions (\emph{e.g.}, noisy measurements and multidimensional signals.)

To start, we observe that $\vy$ is the sum of $K$ exponentials: its $n$th entry is of the form
\begin{equation}\label{eq:soe_full}
y_n= \frac{1}{\sqrt{N}} \sum_{k=0}^{K-1} c_{m_k} \, e^{j 2\pi m_k n / N},
\end{equation}
where $m_k$ is the index\footnote{In this paper we use a zero-based indexing scheme. So the first element of $\vc$ is assigned the index $0$.} of the $k$th nonzero element of $\vc$, and $c_{m_k}$ is the corresponding weight. Writing $\alpha_k \bydef c_{m_k} / \sqrt{N}$ and $u_k \bydef e^{j 2\pi m_k / N}$, we can simplify \eref{soe_full} as
\begin{equation}\label{eq:soe}
y_n=\sum_{k=0}^{K-1} \alpha_k u_k^n.
\end{equation}
Assuming that $K$ is known, we aim to retrieve the coefficients $\set{\alpha_k}$ and the exponentials $\set{u_k}$ from $2K$ consecutive elements $\set{y_n:\ \l \le n < \l + 2K}$. The original $K$-sparse vector $\vc$ can then be reconstructed from $\set{\alpha_k}$ and $\set{u_k}$.

%

The key to Prony's method is a clever use of the algebraic structure of the expression in \eref{soe}. Let
\begin{equation}
\label{eq:PronyPoly}
P(x) = \prod_{k=1}^K (x - u_k) = x^K + h_1 x^{K-1} + h_2 x^{K-2} + \ldots + h_{K-1} x + h_K
\end{equation}
be a $K$th order polynomial whose roots are $\set{u_k}$. Then, it is easy to verify that
\[
y_{n+K} + h_1 \, y_{n+K-1} + h_2 \, y_{n+K-2} + \ldots + h_K \, y_{n} = \sum_{1 \le k \le K} \alpha_k u_k^n P(u_k) = 0.
\]
Writing this identity in matrix-vector form for all indices $n$ such that $\l \le n < \l + K$, we get
\begin{equation}\label{eq:Th}
\vzero = \begin{bmatrix}
y_{\l+K} & y_{\l+K-1} & \cdots & y_\l \\
y_{\l+K+1} & y_{\l+K} & \cdots & y_{\l+1} \\
\vdots & \ddots & \ddots & \vdots \\
y_{\l+2K-2} & \ddots & \ddots & \vdots \\
y_{\l+2K-1} & y_{\l+2K-2} & \cdots & y_{\l+K-1}
\end{bmatrix}
\begin{bmatrix}
1\\
h_1 \\
h_2 \\
\vdots \\
h_K
\end{bmatrix}
\bydef \mT_{K,\l} \vh,
\end{equation}
where, by construction, $\mT_{K,\l}$ is a Toeplitz matrix of size $K\times(K+1)$.

The above equation reveals that the vector of polynomial coefficients $\vh=[1, h_1, . . .,h_K]^T$ is in the null space of $\mT_{K,\l}$. In fact, this condition is sufficient to uniquely identify $\vh$, as guaranteed by the following proposition.
\begin{proposition}\label{prop:rank}
Suppose that $\alpha_k \neq 0$ for all $k$ and that the $K$ parameters $\set{u_k}$ are distinct. Then
\begin{equation}\label{eq:full_rank}
\text{rank}\ \mT_{K,\l} = K.
\end{equation}
\end{proposition}
\begin{IEEEproof}
See, \emph{e.g.}, \cite[Appendix~B]{HormatiRLM:10}.
\end{IEEEproof}

Since $\mT_{K,\l}$ has full row rank, its null space is of dimension one. We can therefore conclude that the vector $\vh$ is the unique vector satisfying the identity \eref{Th}.

In light of the above derivations, we summarize Prony's method as follows:
\begin{enumerate}
\item[(1)] Given the input $y_n$, build the Toeplitz matrix $\mT_{K,\l}$ as in \eref{Th} and solve for $\vh$. This can be achieved by taking the SVD of $\mT_{K,\l}$ and choosing as $\vh$ the (scaled) right-singular vector associated with the zero singular value. The scaling is done so that the first element of $\vh$ is equal to $1$.
\item[(2)] Find the roots of $P(x)=1 + \sum_{n=1}^{K} h_k x^{K-k}$. These roots are exactly the exponentials $\{u_k \}_{k=0}^{K-1}$.
\item[(3)] Given the parameters $\{u_k \}_{k=0}^{K-1}$, find the corresponding weights $\{ \alpha_k \}_{k=0}^{K-1}$ by solving $K$ linear equations as given in \eref{soe}. This is a Vandermonde system of equations which yields a unique solution for the weights $\{ \alpha_k \}_{k=0}^{K-1}$  since $\{ u_k \}_{k=0}^{K-1}$ are distinct.
\end{enumerate}
\begin{remark}
Building the Toeplitz matrix $\mT_{K,\l}$ in \eref{Th} requires $2K$ elements $\set{y_n: \l \le n < \l + 2K}$. Therefore, Prony's method allows us to reconstruct $\set{\alpha_k, u_k}$ and, equivalently, the $K$-sparse vector $\vc$ from \emph{any} $2K$ \emph{consecutive} elements of $\vy$. Moreover, due to the periodicity of the Fourier matrix, these elements of $\vy$ just need to have indices that are consecutive modulo ($N$). For example, $\set{y_{N-2}, y_{N-1}, y_0, \ldots, y_{2K-3}}$ is also a valid choice.
\end{remark}

%



\begin{remark}\label{rem:sparsity}
We have assumed in the above discussions that the sparsity level $K$ is known. In fact, to apply Prony's method, we just need to know an upper bound on the sparsity level. To see this, assume that the true sparsity of $\vc$ is $\widetilde{K}$, for some unknown $\widetilde{K} < K$. Following the same steps in the proof of Proposition~\ref{prop:rank}, we can show that the Toeplitz matrix $\mT_{K, \l}$ in this case is rank-deficient and that its rank is equal to $\widetilde{K}$. Therefore, checking the rank of $\mT_{K, \l}$ allows us to obtain the true sparsity level.
\end{remark}

%
%
\section{Finding Sparse Representations in Fourier and Canonical Bases}
\label{sec:ProSparse}
\subsection{ProSparse: a Polynomial Complexity Algorithm}
We now return to our original problem of finding the sparse representation of a signal in a pair of bases. The observed signal is $\vy= [\mF, \mI] \vx$, where $\mF$ and $\mI$ are two orthogonal matrices corresponding to the Fourier and canonical bases, respectively. We want to retrieve $\vx$ from $\vy$, knowing that $\vx$ has a small number of nonzero elements. 


We begin by noting that the problem is trivial when $\vy$ is made only of spikes, \emph{i.e.}, when the first $N$ entries of $\vx$ are exactly zero. In this case, we can directly retrieve $\vx$ by observing the support set of $\vy$. Likewise, by examining the support of the Fourier transform of $\vy$, we can trivially retrieve the sparse representation of $\vy$ when it is made only of Fourier atoms. Let us assume now that $\vy$ is made of a combination of $K_p$ Fourier atoms and $K_q$ spikes, for some $K_p, K_q \ge 1$. The total sparsity is then defined as $K = K_p + K_q$.


Our proposed algorithm on sparse representation is based on a simple idea: The observation $\vy$ is a mixture of Fourier atoms and spikes, the latter of which are \emph{local}. If we can find an interval of $2K_p$ consecutive entries of $\vy$ that are only due to the Fourier atoms, we can then apply Prony's method presented in \sref{Prony} on these entries to retrieve the $K_p$ Fourier atoms. Once this has been achieved, the spikes can be obtained by removing from $\vy$ the contribution due to the Fourier atoms. Moreover, when both $K_p$ and $K_q$ are small, such ``nice'' intervals should always exist and there might even be a large number of them. 


To quantify the above intuition, denote by $0 \le n_1 < n_2 < \ldots < n_{K_q} \le N-1$ the set of indices corresponding to the $K_q$ spikes. We can count the number of all length-$2K_p$ intervals that are not ``corrupted'' by these spikes as
\begin{equation}\label{eq:number_intervals}
\mathcal{N}(n_1, n_2, \ldots, n_{K_q}) \bydef \#\Big\{\l: 0 \le \l < N \text{ and } \set{\l, \l+1, \ldots,\l+2K_p-1} \cap \set{n_1, n_2, \ldots, n_{K_q}} = \emptyset\Big\}.
\end{equation}
Note that, due to the periodicity of the Fourier exponential $e^{j 2 \pi n/N}$, we should view indices through the modulo (by $N$) operator. This means that $N = 0 \ (\text{mod } N)$ and thus the entry $n=N-1$ is immediately followed by the entry $n=0$, and so on.

\begin{lemma}\label{lemma:bound_P}
Let $\vy$ be a mixture of $K_p$ Fourier atoms and $K_q$ spikes, for some $K_p, K_q \ge 1$. Then, for any choice of spike locations $0 \le n_1 < n_2 < \ldots < n_{K_q} \le N-1$,
\begin{equation}\label{eq:bound_P}
\mathcal{N}(n_1, n_2, \ldots, n_{K_q}) \ge N-2 K_p K_q.
\end{equation}
\end{lemma}
\begin{IEEEproof}
Let $d_i$, for  $1 \le i \le K_q$, denote the number of consecutive entries of $\vy$ that are ``sandwiched'' between two neighboring spikes $n_i$ and $n_{i+1}$. (Here, $n_{K_q+1}$ is defined to be equal to $n_1$.) Then $d_i = (n_{i+1} - n_{i})\ (\text{mod } N)-1$. Clearly, $d_i \ge 0$, and
\begin{equation}\label{eq:sum_d}
\sum_{1 \le i \le K_q} d_i = N - K_q.
\end{equation}
By construction, each of these $K_q$ intervals are ``uncorrupted'' by the spikes. For the $i$th interval, if its length $d_i < 2K_p $, then that particular interval does not contain enough entries for building the Toeplitz matrix in \eref{Th} as required in Prony's method; if however, $d_i \ge 2K_p $, then we can find $d_i - 2 K_p + 1$ (overlapping) subintervals, each of length $2K_p$. It follows that the quantity in \eref{number_intervals} can be computed as
\begin{align}
\mathcal{N}(n_1, n_2, \ldots, n_{K_q}) &= \sum_{1 \le i \le K_q} \max\set{0, d_i - 2K_p +1}\nonumber\\
					  &\ge \sum_{1 \le i \le K_q} (d_i - 2K_p +1)\nonumber\\
					  &= \bigg(\sum_{1 \le i \le K_q}d_i\bigg) - K_q(2K_p - 1).\label{eq:P1}
\end{align}
Substituting \eref{sum_d} into \eref{P1} leads to the bound \eref{bound_P}.     
\end{IEEEproof}

\begin{remark}
Lemma~\ref{lemma:bound_P} guarantees that $\mathcal{N}(n_1, n_2, \ldots, n_{K_q}) \ge 1$, \emph{i.e.}, at least one interval of $2K_p$ consecutive entries containing only Fourier atoms exists,
when $K_p K_q < N/2$. This bound is also tight: Suppose that $K_q$ divides $N$. Let the $K_q$ spikes be evenly spaced to form a ``picket-fence'' signal. In this case, we can have at most $N/K_q - 1$ consecutive entries of $\vy$ that contain only Fourier atoms, before running into another spike. If $K_p K_q \ge N / 2$, the length of such ``clean'' intervals will be strictly smaller than $2 K_p$.
\end{remark}

Based on the above analysis, we are now able to state the following result:
\begin{proposition}\label{prop:ProSparse}
Assume $\mD = [\mF, \mI]$, where $\mF$ and $\mI$ are, respectively, the $N\times N$ Fourier and identity matrices. Let $\vy \in \C^N$ be an arbitrary signal. There exists an algorithm, with a worst-case complexity of $\mathcal{O}(N^{3})$, that finds \emph{all} $(K_p, K_q)$-sparse signals $\vx$ such that
\begin{equation}\label{eq:prosparse_exact}
\vy = \mD \vx \ \mbox{ and } \  K_p K_q < N / 2.
\end{equation}
\end{proposition}
\begin{IEEEproof}
We provide a constructive proof of this proposition by introducing the ProSparse algorithm. We will show that ProSparse finds all $\vx$ satisfying \eref{prosparse_exact}, with the additional constraint that $K_p \le K_q$. The remaining cases, \emph{i.e.}, those $\vx$ satisfying \eref{prosparse_exact} but with $K_p > K_q$, can be obtained through the duality of the Fourier and canonical bases: Suppose that a signal $\vy = [\mF, \mI] \vx$ is made of $K_p$ Fourier atoms and $K_q$ spikes. Denoting by $(\cdot)^\ast$ and $\overline{(\cdot)}$ the Hermitian and complex conjugate operators, respectively, we can then easily verify that a ``dual signal'', $\overline{\mF^* \vy} = [\mI, \mF] \, \overline{\vx}$, is made of $K_q$ Fourier atoms and $K_p$ spikes. Consequently, to recover all $\vx$ satisfying \eref{prosparse_exact}, we just need to run ProSparse twice, with $\vy$ and $\overline{\mF^* \vy}$ being the input each time.
\begin{algorithm}[t]
	\caption{\emph{ProSparse}---Prony's based sparsity}
	\label{alg:ProSparse}
	\algsetup{indent=2em}
	 \vspace{1ex}	
	\begin{algorithmic}
		\REQUIRE A dictionary $\mD = [\mF, \mI]$ and an observed vector $\vy \in \C^N$.
		\ENSURE A set $\mathcal{S}$, containing all $(K_p, K_q)$-sparse signal $\vx$ that satisfies \eref{prosparse_exact}, with $K_p \le K_q$.
		\STATE Initialize $\mathcal{S} = \set{[\vzero^T, \vy^T]^T}$. This is a trivial solution, corresponding to $K_p = 0$ and $K_q = \norm{\vy}_0$.
		\FOR{$K_p =1,2,. . .,\big\lceil \sqrt{N/2} -1 \big\rceil$} 
			\FOR{$\l=0,1,. . .,N-1$}
				\STATE Build the Toeplitz matrix $\mT_{K_p,\l}$ as in \eref{Th}.
	      
	      			\STATE Apply Prony's method on $\mT_{K_p,\l}$ to find the parameters  $\{ \alpha_k, u_k \}$, where $0 \le  k < K_p$.
				
				\IF{$\set{u_k}$ contains $K$ distinct values, with each $u_k \in \set{e^{j 2\pi m / N}: m \in \Z}$}
					\STATE Compute the estimated Fourier contribution $\widehat{y}_n=\sum_{k=0}^{K_p-1}\alpha_ku_k^n$, for $0 \le n < N$.
	     				\STATE Compute the residual $\vr=\vy-{\widehat{\vy}}$ and let $K_q = \norm{\vr}_0$.
	      				\IF{$K_p \le K_q$ and $K_p K_q < N / 2$}
						\STATE Obtain the sparse signal $\vx$ from the Fourier contribution $\widehat{\vy}$ and the residue $\vr$.
						\STATE $\mathcal{S} \Leftarrow \mathcal{S} \cup \set{\vx}$.
					\ENDIF
				\ENDIF
			\ENDFOR
		\ENDFOR
	\end{algorithmic}
\end{algorithm}

Next, we present ProSparse and verify that it indeed has the stated properties. The algorithm, summarized in the insert, operates as follows:  Let $\mathcal{S}$ be the set of solutions the algorithm will return. After initializing $\mathcal{S}$ with the trivial solution that $\vx = [\vzero^T, \vy^T]^T$ (corresponding to $K_p = 0$ and $K_q = \norm{\vy}_0$), set $K_p = 1$. For each $\l=0,1,..,N-1$, build the Toeplitz matrix $\mT_{K_p,\l}$ using a sliding window $[y_\l, y_{\l+1}, \ldots, y_{\l+2K_p - 1}]$ of size $2 K_p$. Apply Prony's method on the $K_p \times (K_p +1)$ Toeplitz matrix $\mT_{K_p,\l}$ in order to retrieve the $K_p$ potential locations $\{ u_k \}$ and amplitudes $\{ \alpha_k \}$ of the Fourier atoms. If the parameters $\set{u_k}$ do not have $K_p$ different values or if they are not in the expected form, \emph{i.e.}, $u_k = e^{j2\pi m/N}$ for some integer $m$, set $\l \Leftarrow \l + 1$ and repeat the process. Otherwise, compute the contribution due to the Fourier atoms as $\widehat{y}_n = \sum_{k=0}^{K_p-1} \alpha_k u_k^n$ for $n=0,1,..N-1$. Remove this contribution from $\vy$ and check whether $K_q$, the number of nonzero entries of the residual, satisfies $K_p K_q <N/2$ and $K_p \le K_q$. If these two conditions are satisfied, use the estimated Fourier atoms and the nonzero entries of the residual as one solution, and add it to the set $\mathcal{S}$. Set $K_p \Leftarrow K_p+1$ and repeat the process up to $K_p=\lceil \sqrt{N/2} -1 \rceil$, where $\lceil c \rceil$ denotes the smallest integer that is greater than or equal to a real number $c$.

By construction of the algorithm, any solution vector $\vx$ in $\mathcal{S}$ must satisfy \eref{prosparse_exact}, subject to the additional condition that $K_p \le K_q$. The opposite direction is also true, \emph{i.e.}, $\mathcal{S}$ contains all such vectors. To see this, we first note that the two constraints, $K_p K_q < N/2$ and $K_p \le K_q$, imply that $K_p < \sqrt{N/2}$. The trivial case, when $K_p = 0$, leads to a solution $\vx = [\vzero^T, \vy^T]^T$, which is added to $\mathcal{S}$ at the beginning of the algorithm. Now suppose that $\vy$ can be written as a combination of $K_p$ Fourier atoms and $K_q$ spikes, such that $1 \le K_p < \sqrt{N/2}$,  $K_p K_q < N/2$ and $K_p \le K_q$. Such a solution will always be found by ProSparse, because when $K_p K_q < N/2$, we know from Lemma~\ref{lemma:bound_P} that  an interval with $2K_p$ consecutive entries due only to Fourier atoms exists. 
Prony's method will then estimate the correct Fourier atoms from these entries and, in this case, the residual will have $K_q$ nonzero entries, satisfying the required conditions.

Finally, we show that ProSparse has a worst-case complexity of $\mathcal{O}(N^3)$. We note that the algorithm has two nested iterations, over $K_p$ and $\l$, respectively. Within the iterations, we apply Prony's method on a matrix of size $K_p \times (K_p+1)$. Finding the polynomial coefficients $\vh$ as in \eref{Th} through SVD has complexity $\mathcal{O}(K_p^3)$. Polynomial root finding in the algorithm has complexity up to $\mathcal{O}(K_pN)$. This is due to the fact that the correct roots in this case can only have $N$ possible choices in the form of $\set{e^{j2\pi m/N}, m=0,1,..,N-1}$ [see also \eref{soe} and \eref{PronyPoly}]. Therefore, we just need to evaluate $P(x)$ of \eref{PronyPoly} at $x=e^{j2\pi m/N}, 0 \le m < N$, to check if this is really a root of the polynomial, whose degree is up to $K_p$. After Prony's, the steps where the Fourier contribution is re-synthesized and where we compute the residue and check its sparsity have complexity $\mathcal{O}(K_p N)$. Therefore, for any fixed $K_p$ and $\l$, the complexity of the algorithm is $\mathcal{O}(K_p^3 + K_pN)$. Since ProSparse loops over $1 \le K_p \le \big\lceil \sqrt{N/2}-1 \big\rceil$ and $0 \le \l < N$, its overall complexity can thus be estimated as $\sum_{1 \le K_p < \lceil \sqrt{N/2} \rceil} \mathcal{O}(K_p^3 N + K_p N^2) \lesssim \mathcal{O}(N^3)$.
\end{IEEEproof}

\begin{remark}
The reason that we consider $K_p \le K_q$  (by using duality) in the ProSparse algorithm is to reduce the computational complexity.
Note that, in this way, $K_q$ just needs to iterate from $1$ to $\big\lceil \sqrt{N/2} -1 \big\rceil$  leading to an overall complexity of $\mathcal{O}(N^3)$.
Without the constraint $K_p < K_q$, we should consider all $K_p$ up to $N$, and this would yield a higher overall complexity.
\end{remark}

In Proposition~\ref{prop:ProSparse}, the condition for successful sparse recovery, $K_p K_q < N/2$, is given in terms of the individual sparsity levels on the Fourier and canonical bases. It is often convenient to have a condition that only depends on the total sparsity level $K = K_p + K_q$. The following result serves this purpose.
\begin{corollary}\label{cor:NP_hard} 
Assume $\mD = [\mF, \mI]$ and let  $\vy \in \C^N$ be an arbitrary signal. There exists an algorithm, with a worst-case complexity of $\mathcal{O}(N^{3})$, that finds \emph{all} $K$-sparse signals $\vx$ such that $\vy = \mD \vx$ and 
\begin{equation}\label{eq:prosparse_joint}
K < \sqrt{2N}.
\end{equation}
In particular, this implies that, if $\vy = \mD \vx$ for some $K$-sparse signal $\vx$ with $K < \sqrt{N}$, the nonconvex problem $(P_0)$, which is known to admit a unique solution in this case, can be solved by an algorithm with polynomial complexity.
\end{corollary}
\begin{IEEEproof}
For any $K_p, K_q \ge 0$, we have $K = K_p + K_q \ge 2 \sqrt{K_p K_q}$. Using this inequality, we can easily see that \eref{prosparse_joint} poses a more restrictive condition than $K_p K_q < N / 2$, meaning that the former implies the latter. The result then follows from Proposition~\ref{prop:ProSparse}.
\end{IEEEproof}

%
%
%
%
%
%
\subsection{Numerical Validation}

To visualize the results of Proposition~\ref{prop:ProSparse}  and Corollary~\ref{cor:NP_hard}, we refer the reader to \fref{VariousBounds:2}, where we plot the exact ProSparse bound $K_p K_q < N/2$ and its simplified version in \eref{prosparse_joint}. In that same figure, we also show the $(P_0)$ bound \eref{unique_solution} and the BP bound \eref{l1_constraint_final}. It is evident that, compared with $(P_0)$ and BP, ProSparse provides performance guarantees over a much wider range of sparsity levels. In what follows, we further validate these theoretical results by presenting two numerical examples where the $(P_0)$ or BP formulation fails to retrieve the original sparse vector while ProSparse remains effective.

\begin{figure}[t]
	\centering
	\subfigure[BP solution]{\label{fig:BP_ProSparse:1}
		\centering
		\includegraphics[width=0.30\textwidth]{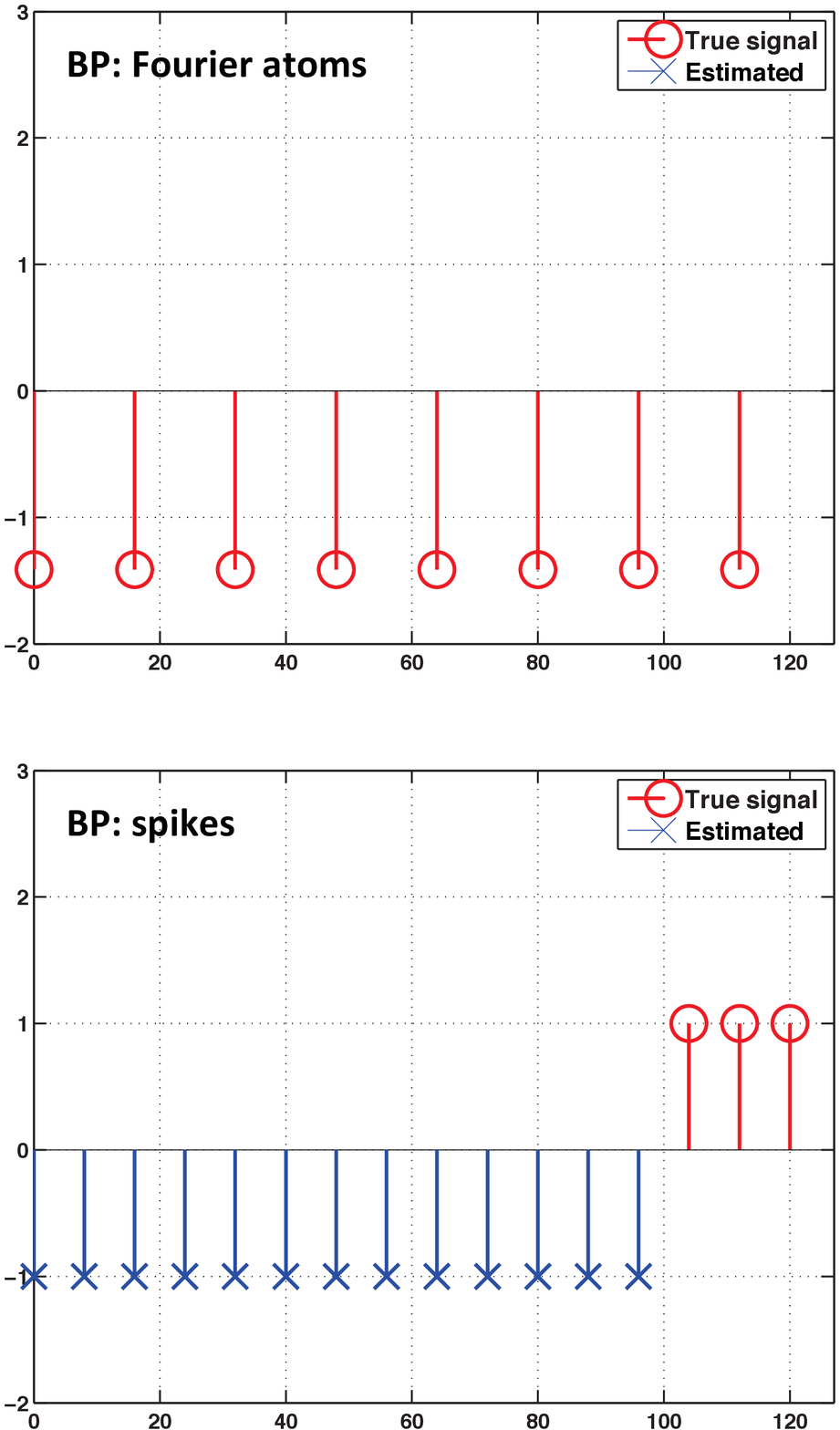}
	}
	\hspace{8ex}
	\subfigure[ProSparse solution]{\label{fig:BP_ProSparse:2}
		\centering
		\includegraphics[width=0.30\textwidth]{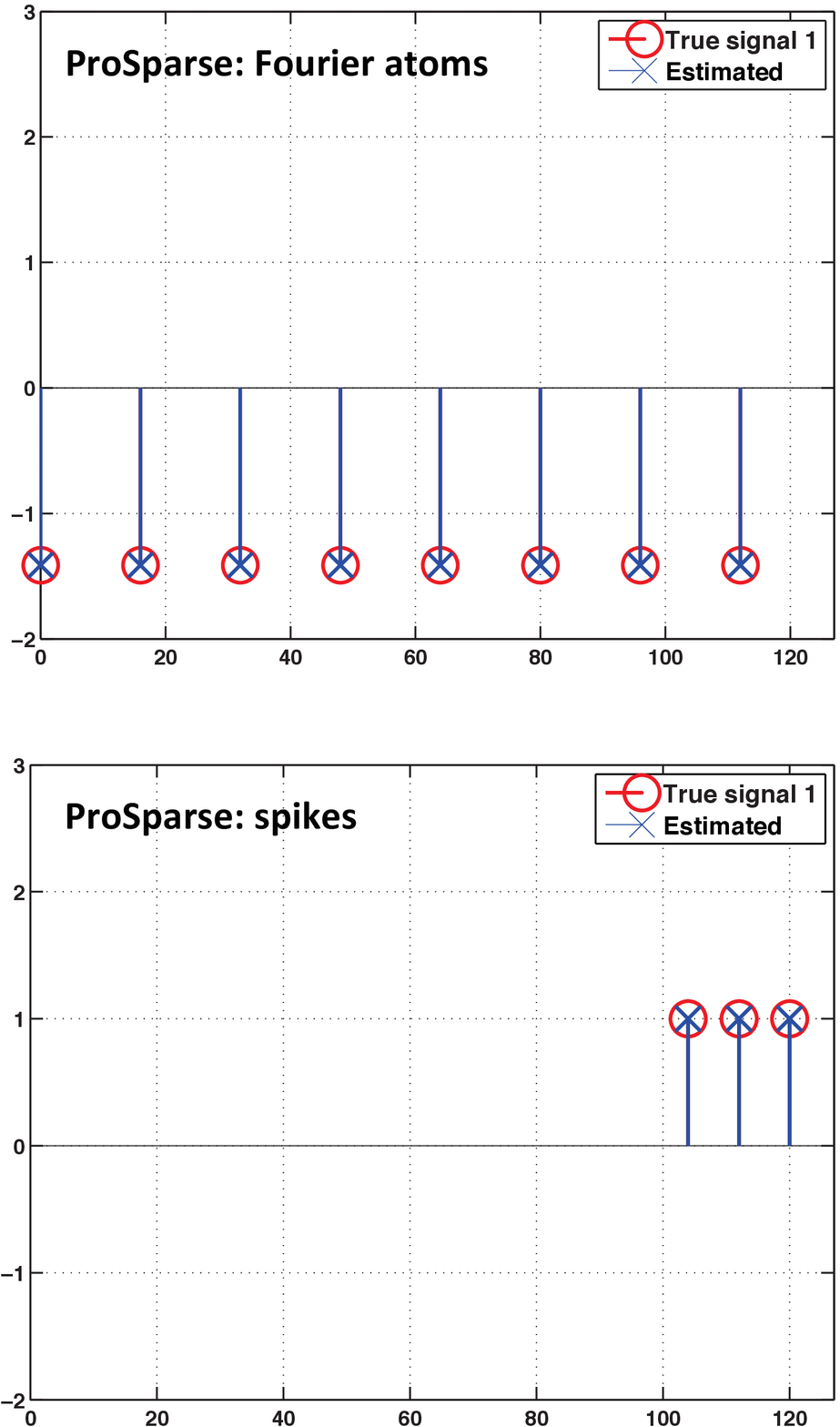}
	}
	\caption{ProSparse vs BP. The observed signal $\vy$ is of length $N = 128$, made of $K_p=8$ Fourier atoms and $K_q=3$ spikes. For comparisons, both the reconstructed signals (in blue) and the ground truth signals (in red) are shown in the figures. (a): The Fourier atoms and spikes recovered by BP. In this case, BP does not find the correct sparse representation. In particular, none of the Fourier atoms has been recovered. (b): ProSparse perfectly retrieves the Fourier atoms and spikes from $\vy$. See Appendix~\ref{appendix:counter} for details on how this example is constructed.}\label{fig:BP_ProSparse}
\end{figure}

\begin{example}[Beyond the BP bound]
In \fref{BP_ProSparse}, we show the results of applying BP and ProSparse, respectively, to find the sparse representation of a signal $\vy$. The length of $\vy$ is $N = 128$, and it is made of $K_p=8$ Fourier atoms and $K_q=3$ spikes. This example has been constructed by adapting the methodology proposed in~\cite{FeuerN:03} and our construction is explained in more details in Appendix~\ref{appendix:counter}.

We note that the sparsity levels are such that the uniqueness condition \eref{unique_solution} for $(P_0)$ holds but the BP bound \eref{tight_bound} is not satisfied. \fref{BP_ProSparse:1} shows the reconstruction results by using BP. In this case, BP fails to find the original sparse representation. In comparison, ProSparse retrieves the correct Fourier atoms and spikes from $\vy$, as shown in \fref{BP_ProSparse:2}. That ProSparse works is expected, since the sparsity levels in this case stay well-within the ProSparse bound ($K_p K_q < N/ 2$) for successful recovery.
\end{example}

\begin{figure}[t]
	\centering
	\subfigure[BP solution]{\label{fig:multiple_solutions:1}
		\centering
		\includegraphics[width=0.30\textwidth]{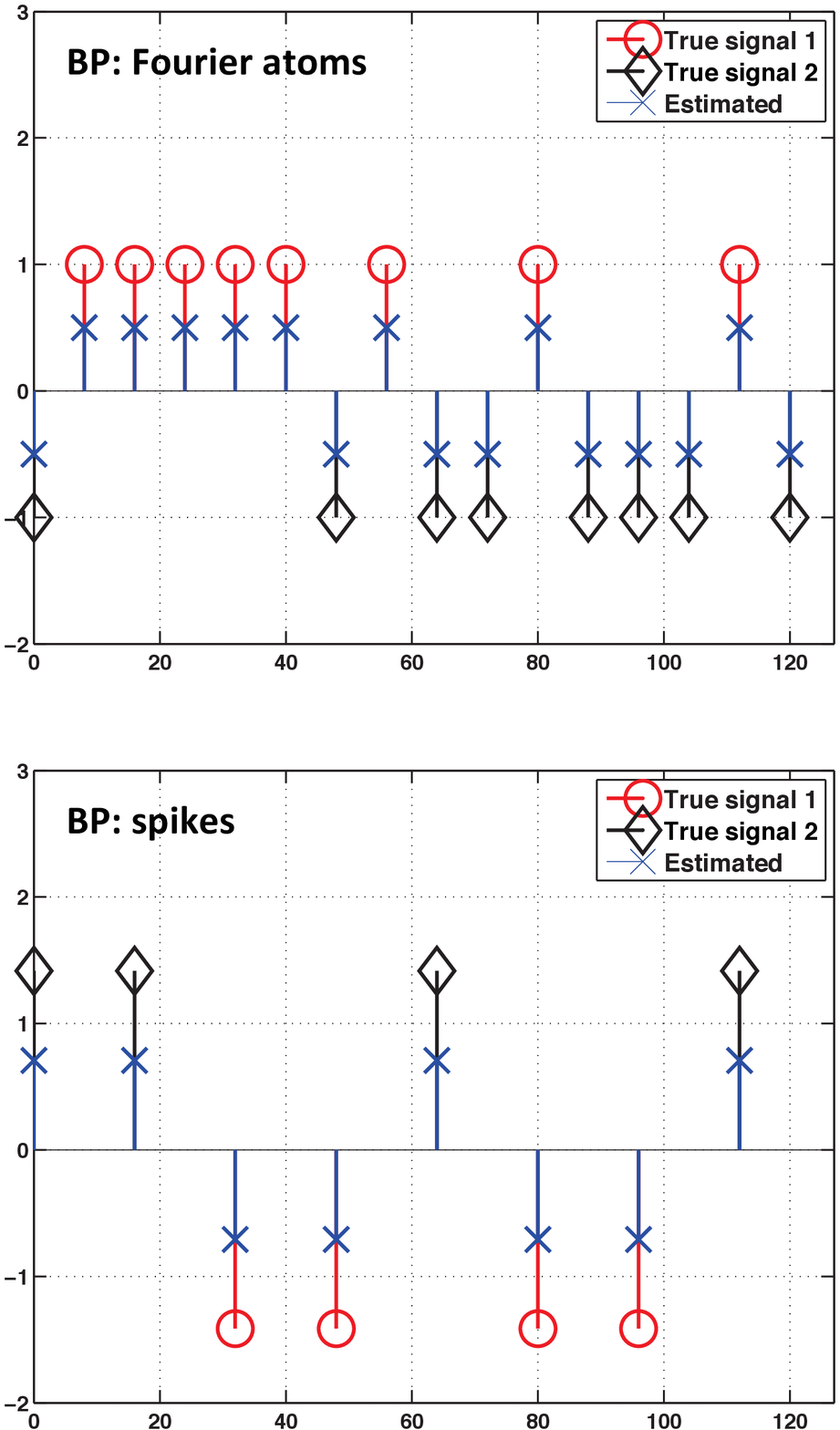}
	}
	\subfigure[ProSparse solution 1]{\label{fig:multiple_solutions:2}
		\centering
		\includegraphics[width=0.30\textwidth]{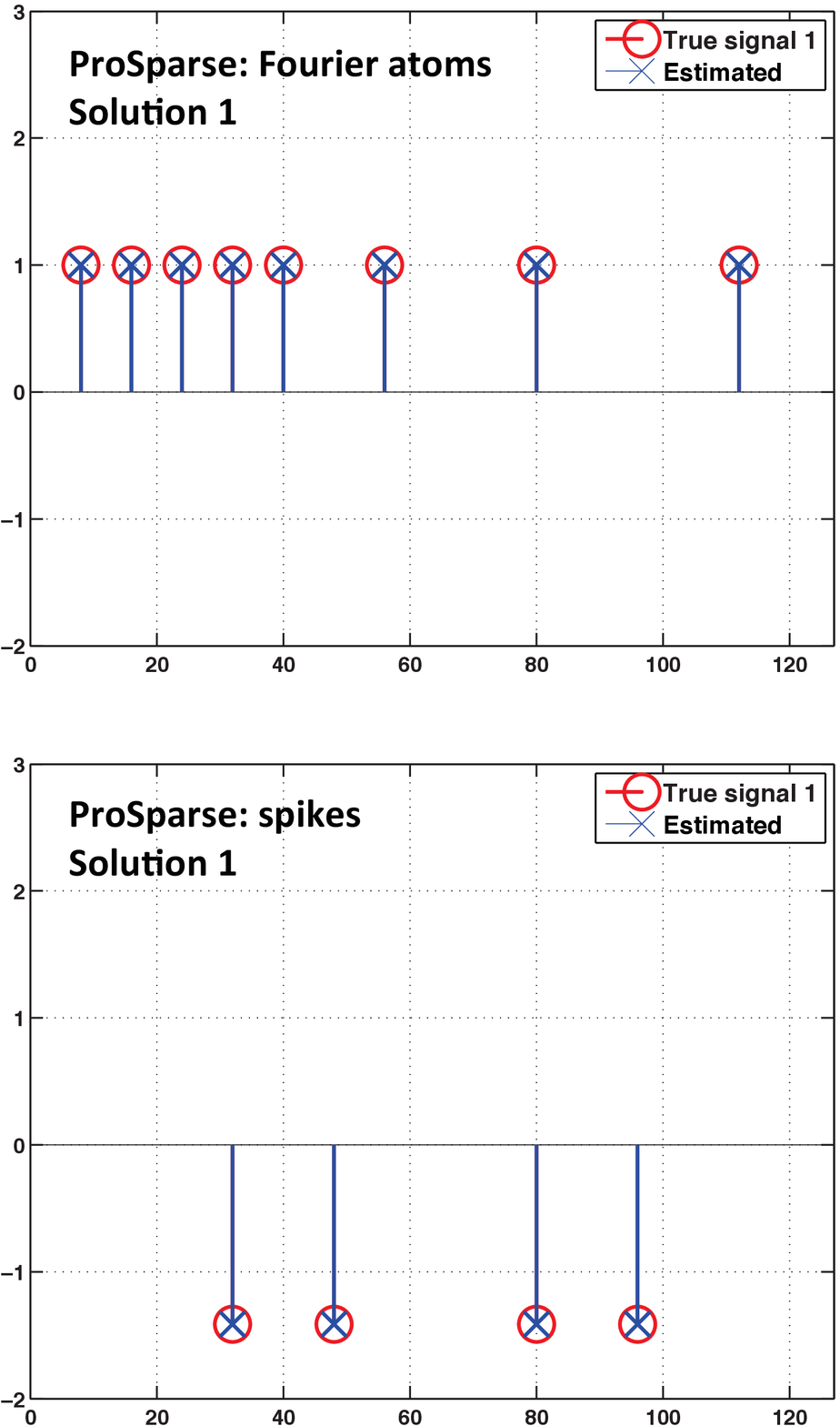}
	}
	\subfigure[ProSparse solution 2]{\label{fig:multiple_solutions:3}
		\centering
		\includegraphics[width=0.30\textwidth]{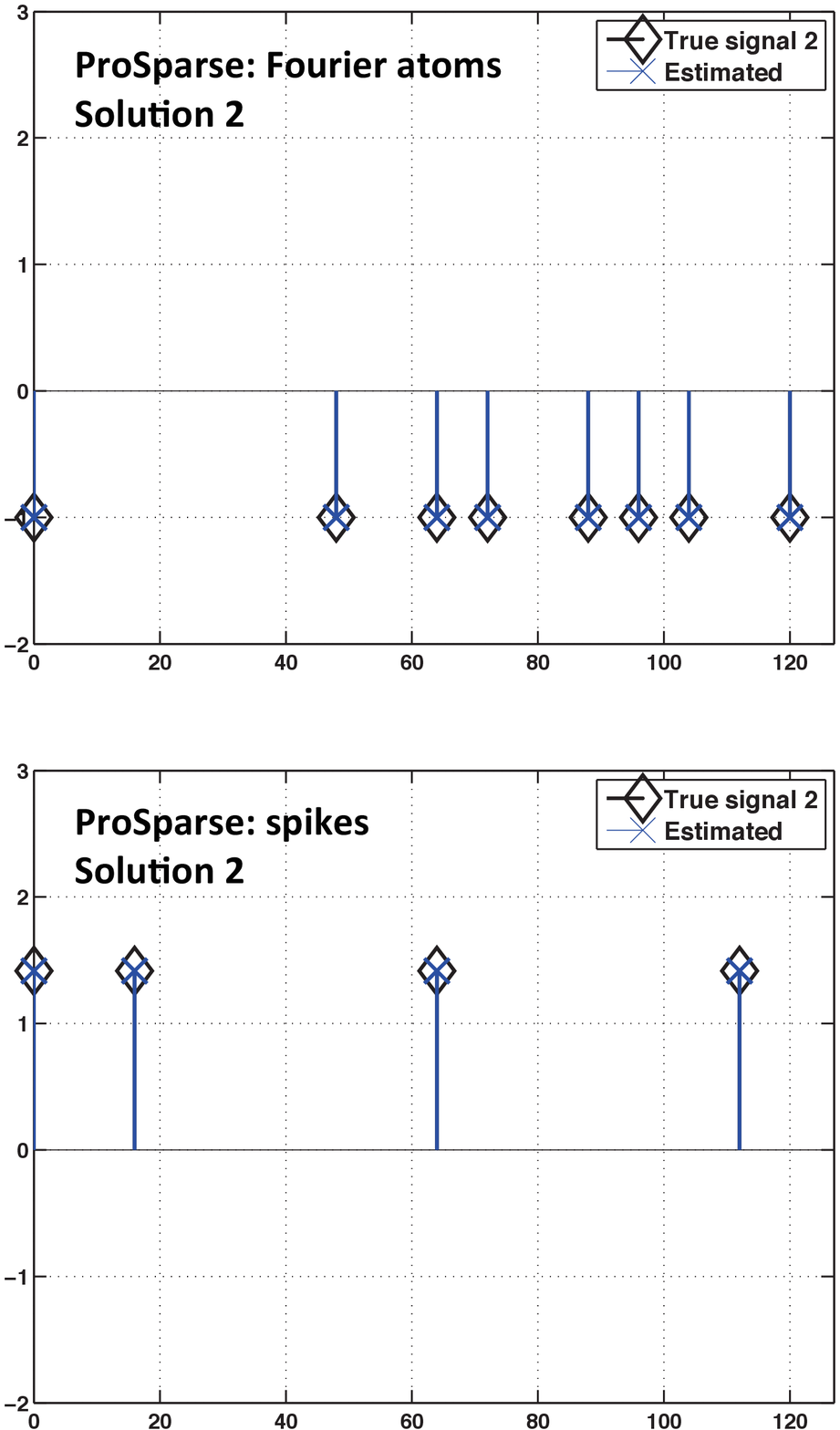}
	}
	\caption{A case when the constraint $\vy = \mD \vx$ admits two $K$-sparse solutions. The observed signal $\vy$ is of length $N = 128$ and the sparsity level is $K = 12$. See Appendix~\ref{appendix:multiple_solutions} for details. In all the figures, the reconstructed signals are shown in blue, whereas the two ground truth signals are shown in red and black, respectively. (a): The BP approach fails to retrieve either of the two solutions. (b) and (c): ProSparse retrieves the two sparse solutions exactly.}
\end{figure}

\begin{example}[Beyond the $(P_0)$ uniqueness bound]
We consider an example where two different $K$-sparse signals lead to the same $\vy$. Clearly, this can be achieved only when $K>\sqrt{N}$, \emph{i.e.}, when $K$ is beyond the $(P_0)$ uniqueness bound. In Appendix~\ref{appendix:multiple_solutions}, we construct one such $\vy$, with parameters $N=128$ and $K=12$. As shown in \fref{multiple_solutions:2} and \fref{multiple_solutions:3}, ProSparse recovers both sparse solutions exactly, whereas the BP approach fails to find either of the two [see \fref{multiple_solutions:1}.]
\end{example}

\section{Generalizations: Other Pairs of Bases}
\label{sec:Extensions}

In this section, we generalize the result of Proposition~\ref{prop:ProSparse} and show that ProSparse-like algorithms can be used to solve the sparse representation problem for a larger family of dictionaries. In what follows, let $\mD = [\mat{\Psi}, \mat{\Phi}]$ be a dictionary consisting of a pair of bases. Unlike in our previous discussions, here we no longer require the two bases to be orthogonal. Let $\vx_p, \vx_q$ be two $N$-dimensional vectors containing $K_p$ and $K_q$ nonzero entries, respectively. Our goal is to recover the $(K_p, K_q)$-sparse vector $\vx = [\vx_p^T, \vx_q^T]^T$ from the measurement $\vy = \mD \vx$.

We note that the ProSparse algorithm presented in \sref{ProSparse} utilizes two fundamental properties of the Fourier and identity matrices: First, each column of the identity matrix has only one nonzero entry so that  $\mI \vx_q$ leaves only a sparse ``footprint'' on the observation vector $\vy$. Most entries of $\vy$ are solely due to the Fourier component $\mF \vx_p$. Second, the algebraic structure of the Fourier matrix allows us to reconstruct the sparse vector $\vx_p$ from only a small number of consecutive entries of $\mF \vx_p$.

We first generalize $\mat{\Phi}$ from the canonical basis to \emph{local bases}. For our purpose, we define the \emph{support length} of a vector $v$ as $\ell(\vv) \bydef \max\set{n: v_n \neq 0} - \min\set{n:v_n \neq 0} + 1$. Essentially, $\ell(\vv)$ is the length of the shortest continuous interval that can cover the support of $\vv$, and it holds that $\ell(\vv) \ge \norm{\vv}_0$.  We call $\mat{\Phi}$ a local basis if all the basis vectors $\set{\mat{\Phi}_i}_{0\le i < N}$ have small support lengths, \emph{i.e.}, the quantity
\begin{equation}\label{eq:max_support_length}
L_{\mat{\Phi}} \bydef \max_i \, \ell(\mat{\Phi}_i)
\end{equation}
is small. For example, when $\mat{\Phi}$ is the canonical basis, we have $L_{\mat{\Phi}} = 1$; When $\mat{\Phi}$ is a banded matrix, $L_{\mat{\Phi}}$ is equal to the bandwidth of that matrix. 

Next, we generalize the Fourier basis $\mat{\Psi}$ to those satisfying the \emph{local sparse reconstruction} property.
\begin{definition}\label{def:local_sparse}
Let $\mat{\Psi}$ be a basis and $\vz = \mat{\Psi} \vc$ for some $K$-sparse vector $\vc$. The basis $\mat{\Psi}$ is said to satisfy the local sparse reconstruction property, if there exists a \emph{polynomial complexity} algorithm that can reconstruct $\vc$ from any $S_{\mat{\Psi}}(K)$ consecutive entries
\begin{equation}\label{eq:consecutive}
\set{z_n, z_{n+1}, \ldots, z_{n+S_{\mat{\Psi}}(K)-1}},
\end{equation}
where $S_{\mat{\Psi}}(K)$ is the minimum number of measurements required at the sparsity level $K$. In what follows, we shall refer to $S_{\mat{\Psi}}(K)$ as the \emph{sparse sampling factor} of $\mat{\Psi}$.
\end{definition}

From our previous discussions, we know that $S_{\mat{\Psi}}(K) = 2K$ for Fourier matrices, and the reconstruction can be done by Prony's method. In Appendix~\ref{appendix:Prony}, we present a more general family of matrices, characterized by 
\begin{equation}\label{eq:Prony_general}
\mat{\Psi} = \mat{\Lambda} \mV \mB,
\end{equation}
where $\mat{\Lambda} \in \C^{N \times N}$ is a diagonal matrix, $\mV \in \C^{N \times M}$ is a Vandermonde matrix with $M \ge N$, and $\mB \in \C^{M \times N}$ is a matrix whose columns have sparse supports. There, we show that, under mild additional conditions on $\mV$ and $\mB$, Prony's method can be used to recover a sparse vector $\vc$ from $S_{\mat{\Psi}}(K) = 2DK$ number of consecutive entries of $\vy = \mat{\Psi} \vc$, where $D$ is some positive integer (see Proposition~\ref{prop:Prony_general} for details.) In particular, it is shown that the DCT matrix can be written in the form of \eref{Prony_general} and that, in this case, we can reconstruct a $K$-sparse vector $\vc$ from any $S_{\mat{\Psi}}(K) = 4K$ consecutive entries of $\vy$.

To state our next result, we need to distinguish two cases: For those matrices (\emph{e.g.}, the Fourier matrix) that have periodic rows, the indices in \eref{consecutive} should be viewed through the modulo (by $N$) operator. In this case, the starting index $n$ can be arbitrarily chosen from $[0, N-1]$, and thus there is a total of $N$ intervals in the form of \eref{consecutive}. However, general basis matrices, such as those characterized in Appendix~\ref{appendix:Prony}, do not have the periodic property. Consequently, the starting index $n$ in \eref{consecutive} can only be chosen from a smaller set, \emph{i.e.}, $[0, N-S_{\mat{\Psi}}(K)]$. In what follows, we refer to matrices in the former case as periodic matrices.

\begin{proposition}\label{prop:other_bases}
Let $\mat{\Phi}$ be a local basis with a maximum support length $L_{\mat{\Phi}}$ as defined in \eref{max_support_length}, and $\mat{\Psi}$ be a basis satisfying the local sparse reconstruction property with a sparse sampling factor $S_{\mat{\Psi}}(K)$ as given in Definition~\ref{def:local_sparse}. Assume $\mD = [\mat{\Psi}, \mat{\Phi}]$ and let $\vy \in \C^N$ be an arbitrary signal. There exists a polynomial complexity algorithm that finds \emph{all} $(K_p, K_q)$-sparse signals $\vx$ such that $\vy = \mD \vx$ and 
\begin{equation}\label{eq:other_bases}
\left(S_{\mat{\Psi}}(K_p) + L_{\mat{\Phi}} - 1\right)(K_q + \tau) < N + \tau L_{\mat{\Phi}},
\end{equation}
where $\tau = 0$ if $\mat{\Psi}$ is a periodic matrix and $\tau = 1$ otherwise.
\end{proposition}
\begin{IEEEproof}
See Appendix~\ref{appendix:other_bases}, where we provide a constructive proof by presenting a generalized version of the ProSparse algorithm.
\end{IEEEproof}

\begin{example}[Fourier and Local Fourier Bases]
Let us assume that $\mD$ is the union of the Fourier basis $\mF_N$ and the local Fourier basis $\mH$, defined as
\[
\mH = 
\left[
\begin{array}{cccc}
\mF_{L} & \vzero & \ldots & \vzero\\
\vzero & \mF_{L} & \ldots & \vzero\\
\vdots & \vdots & \ddots & \vdots\\
\vzero & \ldots & \vzero & \mF_L
\end{array}
\right],
\]
where the subscripts in $\mF_N$ and $\mF_L$ indicate that they are the Fourier matrices of size $N \times N$ and $L \times L$, respectively. Note that when $L = 2$, the matrix $\mH$ can also be seen as the Haar wavelet basis with one level of decomposition. The mutual coherence of the dictionary is $\mu(\mD)=\sqrt{L/N}$, and thus, from \eref{P_0_mutual}, the uniqueness condition for $(P_0)$ is 
\[
K_p + K_q < \sqrt{N/L}.
\]
To apply the result of Proposition~\ref{prop:other_bases}, we substitute $S_{\mat{\Psi}}(K_p) = 2 K_p$, $\tau = 0$, and $L_{\mat{\Phi}} = L$ into \eref{other_bases} and get the ProSparse bound as
\begin{equation}\label{eq:bound_Fourier_local_Fourier}
(2 K_p + L - 1) K_q < N.
\end{equation}
For an easier comparison between the above two bounds, we can verify that a sufficient condition for \eref{bound_Fourier_local_Fourier} to hold is\footnote{Here, we 
have used again the inequality  $x+y \geq 2 \sqrt{xy}$ with $x=K_p+(L-1)/2$ and $y = K_q$.}
\begin{equation}\label{eq:Fourier_local_Fourier_total}
K_p + K_q < \sqrt{2N} - (L-1)/2.
\end{equation}
If we choose, for example, $L = \sqrt{N}$, then the $(P_0)$ problem is unique when the total sparsity is below $N^{1/4}$. In contrast, \eref{Fourier_local_Fourier_total} implies that the generalized ProSparse algorithm can handle a much wider range of sparsity levels, recovering all signals whose sparsity level is below $(\sqrt{2} - 0.5) \sqrt{N}+0.5$.
\end{example}



\begin{example}[DCT and Canonical Bases]
Let $\mD= [\mat{\Psi}, \mI ]$ be the union of the DCT and canonical bases. The mutual coherence in this case is $\mu(\mD)=\sqrt{2/N}$. Consequently, unicity of $(P_0)$ is guaranteed when
\[
K_p + K_q < \sqrt{N/2}.
\]
We have shown in Appendix~\ref{appendix:Prony} that $S_{\mat{\Psi}}(K_p) = 4K_p$ for DCT matrices. Substituting this quantity, together with $\tau = 1$ (since $\mat{\Psi}$ is not periodic) and $L_{\mat{\Phi}} = 1$ into \eref{other_bases}, we conclude that ProSparse can retrieve all $(K_p, K_q)$-sparse signals when $4 K_p (K_q + 1) < N + 1$. A sufficient condition for this bound to hold is
\[
K_p + K_q < \sqrt{N+1} - 1. 
\]
Therefore, in this case, the ProSparse bound is again a superset of the $(P_0)$ bound.
\end{example}

\begin{example}[Random Gaussian and Canonical Bases]
In this example, we consider $\mD = [\mat{\Psi}, \mI]$, where the entries $\set{\Psi_{i,j}}$ of the first basis matrix are independent realizations of Gaussian random variables, \emph{i.e.}, $\Psi_{i, j} \sim \mathcal{N}(0, 1)$. Adapting standard results in compressed sensing \cite{CandesT:06, Donoho:06, BaraniukDDW:08}, we verify in Appendix~\ref{appendix:Gaussian} the following result: Define
\begin{equation}\label{eq:S_Gaussian}
S_{\mat{\Psi}}(K) = \max\set{p(N), \min\set{N, c_1 K \log(N/K)}},
\end{equation}
where $p(N)$ is some positive function of $N$ and $c_1$ is some constant. Then there exist constants $c_1, c_2 > 0$, which do not depend on $N$ or $K$, such that, with probability at least 
\begin{equation}\label{eq:probability}
1 - 2 N^2 e^{-c_2 p(N)}, 
\end{equation}
the random matrix $\mat{\Psi}$ will satisfy the following property: Let $\vc$ be any $K$-sparse vector. One can efficiently reconstruct $\vc$ from any $S_{\mat{\Psi}}(K)$ consecutive (modulo $N$) entries of the vector $\vz = \mat{\Psi} \vc$. By properly choosing $p(N)$, the probability in \eref{probability} can be made arbitrarily close to one for sufficiently large $N$, and thus, the matrix $\mat{\Psi}$ will satisfy the desired property with high probabilities. It then follows from Proposition~\ref{prop:other_bases} that, for those suitable $\mat{\Psi}$, we can find all $(K_p, K_q)$-sparse signals $\vx$ from $\vy = [\mat{\Psi}, \mI] \, \vx$ if
\[
S_{\mat{\Psi}}(K_p) K_q < N,
\]
where $S_{\mat{\Psi}}(\cdot)$ is the function defined in \eref{S_Gaussian}.
\end{example}



Finally, we make the following observation. Denote by $\mD= [\mat{\Psi}, \mat{\Phi}]$ a dictionary for which ProSparse can be used successfully. Namely, $\mD$ can be the union of any pair of bases discussed so far. Let $\mA$ be an arbitrary $N \times N$ invertible matrix. Then, ProSparse can also be used on the dictionary $\widetilde{\mD}= [\mA \mat{\Psi}, \mA \mat{\Phi}]$. This fact can be trivially demonstrated by noting that, given $\vy= \mA \mD \vx$, we can return to the original dictionary by working with $\widetilde{\vy} = \mA^{-1} \vy$.

\section{Conclusions}
\label{sec:Conclusions}

We considered the problem of finding the sparse representations of a signal in the union of two bases. We introduced a new polynomial complexity algorithm {\em ProSparse} which, for the case of Fourier and canonical bases, is able to find all the sparse representations 
of the signal under the condition that $K_p K_q < N / 2$ (or, in terms of the total sparsity level, $K_p + K_q <\sqrt{2N}$.) The new algorithm provides deterministic performance guarantees over a much wider range of sparsity levels than do existing algorithms such as the nonconvex $\ell_0$-minimization or BP. In particular, our results imply that the $\ell_0$-minimization problem for sparse representation is not NP-hard under the unicity condition and when the dictionary is the union of Fourier and canonical bases. Furthermore, we have shown that the proposed algorithm can be extended to other relevant pairs of bases, one of which must have local atoms. Examples include the Fourier and local Fourier bases, the DCT and canonical bases, and the random Gaussian and canonical bases.

\appendix
\section*{}

\subsection{Constructing Counterexamples} 
\label{appendix:counter}

In~\cite{FeuerN:03},  Feuer and Nemirovsky constructed an example showing that $(P_0)$ and $(P_1)$ are not equivalent, for the case where $\mD= [\mH, \mI]$. Here, $\mH$ is the scaled Hadamard matrix, with $\mH^T \mH = \mI$. In this appendix, we summarize the construction in \cite{FeuerN:03} and show how to adapt it to the case where $\mD$ is the union of Fourier and canonical bases.

We first note that  the uniqueness condition \eref{P_0_mutual} for the union of Hadamard and canonical bases is the same as that for Fourier and canonical bases. In both cases, the solution to $(P_0)$ is unique when $K<\sqrt{N}$. In what follows, we set $N=2^{2d-1}$ for some positive integer $d$, and let $K=\lfloor \sqrt{N} \rfloor$.

The key idea behind Feuer and Nemirovsky's construction \cite{FeuerN:03} is to find a vector $\vz \in \R^{2N}$ such that $\mD \vz = \vzero$ and that
\begin{equation}\label{eq:z_condition}
\norm{\vz}_1 = \sum_{n=0}^{N-1} \abs{z_n} < 2 \sum_{k=0}^{K-1} \abs{z_{n_k}},
\end{equation}
where the indices $n_0, n_1, \ldots, n_{K-1}$ correspond to the $K$ largest absolute values of $\vz$. We will provide explicit constructions of $\vz$ a little later. For now, assume that such a vector $\vz$ has already been found.

Given $\vz$, one then builds two vectors $\vx, \widetilde{\vx} \in \mathbb{R}^{2N}$ as follows: a $K$-sparse vector $\vx$ whose nonzero entries satisfy $x_{n_k} = -2 z_{n_k}$, where the indices $\set{n_k}$ are same as those in \eref{z_condition}; and a second vector $\widetilde{\vx}=\vz+\vx$. Given $\vy = \mD \vx$, we know that $\vx$ is the unique solution of $(P_0)$, since the bound $K < \sqrt{N}$ is satisfied here. However, $\vx$ is not the solution of $(P_1)$. To see this, we note that, since $\mD \vz = 0$, we must have $\vy = \mD \vx = \mD \widetilde{\vx}$. Meanwhile, by construction, 
\[
\norm{\widetilde{\vx}}_1 = \norm{\vz}_1 < \norm{\vx}_1,
\]
where the inequality is due to \eref{z_condition}. Consequently, given $\vy$, the solution to $(P_1)$ will not be $\vx$, since there is at least one alternative vector, $\widetilde{\vx}
$, satisfying the same synthesis equation $\vy = \mD \widetilde{\vx}$ but with a smaller $\ell_1$ norm. 

Next, we present explicit constructions of the vector $\vz$ with the desired properties. A suitable $\vz$ was found in \cite{FeuerN:03} for the case of Hadamard and identity matrices. Here, we modify that construction so that it is suitable to the case of Fourier and canonical bases. Recall that $N = 2^{2d-1}$. Define $m=2^d$, and let $\vv \in \R^N$ be a ``picket-fence'' signal, containing exactly $N/m$ uniformly spaced nonzero entries, all of which are equal to $\sqrt{2}$. More precisely, $\vv$ is equal to the following Kronecker product $\sqrt{2} (\vec{1} \otimes \ve_0)$, where $\vec{1} \in \R^{N/m}$ is a vector of all $1$'s, and $\ve_0 = [1, 0, 0, \ldots, 0]^T$ is the first canonical basis vector in $\R^m$. Let
\begin{equation}\label{eq:vec_z}
\vz= 
\begin{bmatrix}
\vv \\
-\mF \vv \\
\end{bmatrix}.
\end{equation}
By construction, $\mD \vz = [\mF, \mI] \vz = \vzero$. Meanwhile, it can be verified that, just like $\vv$, the vector $\mF \vv$ is also a ``picket-fence'' signal, containing $m$ uniformly spaced nonzero entries, all of which are equal to $1$. Since $\vz$ is a concatenation of $\vv$ and $-\mF\vv$, it contains exactly $2^{d-1} + 2^{d}$ nonzero entries, the first $2^{d-1}$ of which are equal to $\sqrt{2}$ and the remaining $2^{d}$ of which are equal to $1$. Now we just need to verify that $\vz$ satisfies \eref{z_condition}. To that end, we first note that $K=\lfloor \sqrt{N} \rfloor = \lfloor 2^{d-0.5} \rfloor \ge 2^{d-1}$. It follows that
\begin{align}
2 \sum_{k=0}^{K-1} \abs{z_{n_k}} - \sum_{n=0}^{N-1} \abs{z_n} &= 2 \left(2^{d-1}\sqrt{2} + \big\lfloor 2^{d-0.5} - 2^{d-1} \big\rfloor \right) - \left(2^{d-1} \sqrt{2} + 2^d\right)\nonumber\\
				&= 2 \left(2^{d-1}\big(\sqrt{2}/2 - 1\big) + \big\lfloor 2^{d-1}(\sqrt{2}-1) \big\rfloor\right).\label{eq:gap}
\end{align}
It is easy to show that, for all $d \ge 4$, the right-hand side of \eref{gap} is strictly positive. Therefore, the vector $\vz$ as constructed above satisfies the condition \eref{z_condition} for all $d \ge 4$.

\subsection{Constructing Examples Where $\vy=\mD \vx$ Admits Two Sparse Solutions} 
\label{appendix:multiple_solutions}

We show how to construct two $K$-sparse vectors $\vx_0, \vx_1 $ such that   $\mD \vx_0=\mD \vx_1$, where $\mD = [\mF, \mI]$. We start with the vector $\vz$ defined in \eref{vec_z} in Appendix~\ref{appendix:counter}. By construction, $\mD \vz=0$ and $\vz$ has exactly $L=2^{d-1}+2^{d}$ nonzero entries, where $d$ is a positive integer satisfying  $2^{2d-1}=N$. We set $d \ge 2$ so that $L$ is even. The two vectors $\vx_1,\vx_0$ are then easily built by assigning $K = L/2$ randomly chosen nonzero entries of $\vz$ to $\vx_0$ and then setting $\vx_1 = \vx_0 - \vz$. Since $\mD \vz = 0$, we must have $ \mD \vx_0 = \mD \vx_1$. Meanwhile, both vectors have the same sparsity level $K = L/2$, which is beyond the $(P_0)$ uniqueness bound \eref{unique_solution} but still within the ProSparse bound given in \eref{prosparse_joint}.

\subsection{Generalizing Prony's Method}
\label{appendix:Prony}

In \sref{Prony}, we showed that Prony's method provides an efficient way to reconstruct a sparse signal $\vc$ from a small number of consecutive entries of the observation 
vector $\vy = \mF \vc$, where $\mF$ is the DFT matrix. Here, we generalize Prony's method for sparse recovery to a larger class of bases, all of which have the following form:
\begin{equation}\label{eq:Prony_general2}
\mat{\Psi} = \mat{\Lambda} \mV \mS,
\end{equation}
where $\mat{\Lambda} \in \C^{N \times N}$ is a diagonal matrix; $\mV \in \C^{N \times M}$ is a Vandermonde matrix whose rows are the powers of a vector $\vp = [p_0, p_1, \ldots, p_{M-1}]$ with \emph{distinct} elements, \emph{i.e.}, $[\mV]_{n, m} = p_m^n$ for $0 \le n < N, 0 \le m < M$; and $\mS \in \C^{M \times N}$ is a matrix whose columns are all sparse. In particular, we assume that, for all $0 \le n < N$, the $n$th column of $\mS$, denoted by $\vs_n$, satisfies
\[
\norm{\vs_n}_{0} \le D,
\]
for some $D > 0$.

\begin{proposition}\label{prop:Prony_general}
Let $\vy = \mat{\Psi} \vc$, where $\vc$ is a $K$-sparse vector and $\mat{\Psi}$ is an invertible matrix in the form of \eref{Prony_general2}. If $\mat{\Lambda}, \mV$ and $\mS$ satisfy the conditions stated above, we can use Prony's method to recover $\vc$ from any $2DK$ consecutive entries of $\vy$.
\end{proposition}
\begin{IEEEproof}
The case when $2DK \ge N$ is trivial: since the entire vector $\vy$ is available, we can reconstruct $\vc$ by a direct linear inversion, \emph{i.e.}, $\vc = \mat{\Psi}^{-1} \vy$. In what follows, we assume that $2DK < N$.  

The basis matrix $\mat{\Psi}$ being invertible implies that the diagonal matrix $\mat{\Lambda} = \text{diag}\set{\lambda_0, \lambda_1, \ldots, \lambda_{N-1}}$ must also be invertible. Introducing two vectors $\vz \bydef\mat{\Lambda}^{-1} \vy$ and $\vx \bydef \mS \vc$, we can simplify the relationship $\vy = \mat{\Lambda} \mV \mS \vc$ as
\[
\vz = \mV \vx.
\]
Since $\vx$ is a linear combination of $K$ vectors, each of which has a sparsity level bounded by $\mD$, we must have $\norm{\vx}_0 \le DK < N/2$. Let $\set{m_0, m_1, \ldots, m_{DK-1}}$ denote indices of the nonzero elements of $\vx$. It follows from the Vandermonde structure of $\mV$ that the $n$th entry of $\vz$ can be written as
\[
z_n = \sum_{k=0}^{DK-1} {x_{m_k}} \, p_{m_k}^n,
\]
which has the same ``sums of exponentials'' form as in \eref{soe}. Consequently, by following the same derivations in \sref{Prony}, we can show that Prony's method\footnote{It is possible that the number of nonzero elements of $\vx$ is less than $DK$. This will not cause a problem for Prony's method, since the algorithm only needs to know an upper bound on the true sparsity level. See Remark~\ref{rem:sparsity} in \sref{Prony} for more details.} can be used to reconstruct $\set{x_{m_k}}$ and $\set{p_{m_k}}$, and therefore $\vx$, from any $2DK$ consecutive entries of $\vz$. Since $z_n = y_n /\lambda_n$, this is equivalent to requiring $2DK$ consecutive entries of $\vy$. Finally, since $\mat{\Psi}$ is invertible, the matrix $\mS$ must necessarily have full column-rank. Thus, the $K$-sparse vector $\vc$ can be obtained from $\vx$ through a simple linear inversion $\vc = (\mS^T \mS)^{-1} \mS^T \vx$.
\end{IEEEproof}

\begin{example}[Discrete Cosine Transform]
Let $\mat{\Psi}$ be the DCT matrix, whose $(n,m)$th entry is
\[
\psi_{n,m}= b(n) \sqrt{\frac{2}{N}}\cos \frac{ \pi n (m+0.5)}{N}, \qquad \text{for } 0 \le n, m, < N
\]
with
\[
b(n)=\left\{\begin{array}{ll}
 1/\sqrt{2} & \mbox{if } n=0, \\
 1 & \mbox{if } 1 \le n < N.
\end{array}
\right.
\]
Using the identity 
$$
2 \cos \frac{ \pi n (m+0.5)}{N}= e^{j\pi n(m+0.5)/N}+e^{-j\pi n(m+0.5)/N},
$$
we can factor $\mat{\Psi}$ in the form of \eref{Prony_general2}: The diagonal matrix is $\mat{\Lambda} = \frac{1}{\sqrt{2N}}\text{diag}\set{b_0, b_1, \ldots, b_{N-1}}$; the Vandermonde matrix $\mV$ is generated by powers of the row vector $\vp = [p_0, p_1, \ldots, p_{2N-1}]^T$, where $p_m = e^{-j\pi(m+0.5)/N}$ for $0 \le m \le N-1$ and $p_m = -e^{j\pi(m+0.5)/N}$ for $N \le m < 2N$; the third matrix $\mS = [1, 1]^T \otimes \mI_N$, where $\otimes$ denotes the matrix Kronecker product and $\mI_N$ is the $N \times N$ identity matrix.

We can easily verify that the entries of $\vp$ are all distinct and that each column of $\mS$ has exactly two nonzero entries (thus, $D = 2$). It follows from Proposition~\ref{prop:Prony_general} that Prony's method is applicable in this case: We can recover a $K$-sparse vector $\vc$ from any $4K$ consecutive entries of $\vy$.
\end{example}

\subsection{Generalized ProSparse Algorithm}
\label{appendix:other_bases}

In this appendix, we provide a constructive proof of Proposition~\ref{prop:other_bases}. To do that, we first need to establish the following result, which is a more general version of Lemma~\ref{lemma:bound_P}. Let $\mat{\Phi}_{n_1}, \mat{\Phi}_{n_2}, \ldots, \mat{\Phi}_{n_{K}}$ be a set of $K$ atoms from the local basis $\mat{\Phi}$. Similar to \eref{number_intervals}, we can count the number of all intervals of length $S$ that are not ``corrupted'' by any of these atoms as 
\begin{equation}\label{eq:number_intervals_general}
\begin{aligned}
&\mathcal{N}_{\mat{\Phi}, \tau}(S; n_1, n_2, \ldots, n_K) \\
&\qquad\bydef \#\Big\{\l: 0 \le \l \le N-1-\tau (S-1) \text{ and } \set{\l, \l+1, \ldots,\l+S-1} \cap \bigcup_{1\le i \le K}\text{supp}\, \mat{\Phi}_{n_i} = \emptyset \Big\},
\end{aligned}
\end{equation}
where $\text{supp}\, \mat{\Phi}_{n_i}$ denotes the support of $\mat{\Phi}_{n_i}$, and $\tau$ is a binary value: $\tau = 0$ if the indices in \eref{number_intervals_general} are periodic on the torus $[0, 1, \ldots, N-1)$, and $\tau = 1$ otherwise.

\begin{lemma}\label{lemma:bound_general}
Let $\mat{\Phi}$ be a local basis with a maximum support length $L_{\mat{\Phi}}$. For any choice of $K$ basis vectors $\set{\mat{\Phi}_{n_i}}_{1\le i \le K}$, it holds that
\begin{equation}\label{eq:bound_general}
\mathcal{N}_{\mat{\Phi}, \tau}(S; n_1, n_2, \ldots, n_K)  \ge N + \tau L_{\mat{\Phi}} - (S + L_{\mat{\Phi}} - 1)(K + \tau).
\end{equation}
\end{lemma}
\begin{IEEEproof}
By the definition of maximum support length \eref{max_support_length}, the support of each basis vector must be fully inside of an interval of length $L_{\mat{\Phi}}$, \emph{i.e.},
\[
\text{supp}\, \mat{\Phi}_{n_i} \subseteq \mathcal{I}_i \bydef [m_i, m_i + 1, \ldots, m_i + L_{\mat{\Phi}} - 1].
\]
Without loss of generality, we assume that the starting indices, $\set{m_i}$, are in ascending order, with $m_1 \le m_2 \le \ldots \le m_K$.

We first consider the case when $\tau = 1$, \emph{i.e.}, the indices are not periodic. 
Let $d_i$ denote the length of the ``uncorrupted'' interval that strictly falls between two neighboring intervals $\mathcal{I}_i$ and $\mathcal{I}_{i+1}$. It is easy to verify that
\begin{equation}\label{eq:d_intervals}
d_i = \max\set{0, m_{i+1} - m_i - L_{\mat{\Phi}}},
\end{equation}
for $0 \le i \le K$. Note that we define $m_0 = -L_{\mat{\Phi}}$ and $m_{K+1} = N$, so that $d_0$ and $d_K$ count, respectively, the number of indices in front of the first interval and the number of those after the last interval. For those intervals with $d_i \ge S$, we can find $d_i - S + 1$ (overlapping) subintervals, each of length $S$. It follows that
\begin{align}
\mathcal{N}_{\mat{\Phi}, \tau=1}(S; n_1, n_2, \ldots, n_K)&\ge \sum_{0 \le i \le K} \max\set{0, d_i - S +1}\label{eq:non_periodic}\\
					  &\ge \sum_{0 \le i \le K} (m_{i+1} - m_i -L_{\mat{\Phi}}) - (K+1)(S-1)\nonumber\\
					  &= m_{K+1} - m_0 - (K+1)(S+L_{\mat{\Phi}}-1),\label{eq:non_periodic2}
\end{align}
which leads to the bound in \eref{bound_general} for $\tau = 1$. 

The proof for the case when $\tau = 0$, \emph{i.e.}, when the indices are periodic, is similar. Setting $m_0 = m_1$ and $m_{K+1} = N+m_1$ in \eref{d_intervals}, we have $d_0 = 0$ and $d_K$ measures the number of indices (modulo $N$) between the last interval $\mathcal{I}_K$ and the first interval $\mathcal{I}_1$. Unlike in \eref{non_periodic} where we sum over $0 \le i \le K$, here, since $d_0 = 0$, we just need to sum over $1 \le i \le K$ and get
\[
\mathcal{N}_{\mat{\Phi}, \tau=0}(S; n_1, n_2, \ldots, n_K) \ge \sum_{1 \le i \le K} \max\set{0, d_i - S +1}.
\]
Following the same steps in reaching \eref{non_periodic2}, we can show that the above inequality yields the bound \eref{bound_general} for $\tau = 0$.
\end{IEEEproof}

\begin{IEEEproof}[Proof of Proposition~\ref{prop:other_bases}]
We provide a constructive proof of Proposition~\ref{prop:other_bases} by presenting in Algorithm~\ref{alg:ProSparse_generalized} a generalized version of ProSparse. We first show that the algorithm can find every $(K_p, K_q)$-sparse signal $\vx = [\vx_1^T, \vx_2^T]^T$ satisfying $\vy = \mD \vx$ and \eref{other_bases}. To see this, we note that, if \eref{other_bases} holds, then Lemma~\ref{lemma:bound_general} guarantees the existence of at least one length-$S_{\mat{\Psi}}(K_p)$ interval that is only due to the atoms in $\mat{\Psi}$. Since the algorithm searches over all possible values of $K_p$ and all possible choices of the intervals, the above-mentioned interval will be examined by the algorithm. By the definition of $S_{\mat{\Psi}}(K_p)$, such an interval is sufficient for us to reconstruct the $K_p$-sparse signal $\vx_1$ with polynomial complexity. Given $\vx_1$, the second half of $\vx$ can then be obtained by removing from $\vy$ the contributions of $\mat{\Psi}$, \emph{i.e.}, $\vx_2=\mat{\Phi}^{-1}(\vy- \mat{\Psi}\vx_1)$.

For computational complexity, we note that the generalized algorithm has two nested iterations, over $1 \le K_p \le N$ and $0\le \l \le \big(N-1-\tau(S_{\mat{\Psi}}(K_p)-1)\big)$, respectively. Within the iterations, the steps in estimating $\vx_1$ and $\vx_2$ both take polynomial time. Therefore, the overall complexity of the algorithm is polynomial in $N$.
\end{IEEEproof}

\begin{algorithm}[t]
	\caption{Generalized ProSparse for Sparse Signal Reconstruction}
	\label{alg:ProSparse_generalized}
	\algsetup{indent=2em}
	 \vspace{1ex}	
	\begin{algorithmic}
		\REQUIRE A dictionary $\mD = [\mat{\Psi}, \mat{\Phi}]$ and an observed vector $\vy \in \C^N$.
		\ENSURE A set $\mathcal{S}$, containing all $(K_p, K_q)$-sparse signal $\vx$ that satisfies the conditions $\vy = \mD \vx$ and \eref{other_bases}.
		\STATE Set $\tau = 0$ if $\mat{\Psi}$ is a periodic matrix and $\tau = 1$ otherwise.
		\STATE Initialize $\mathcal{S} = \set{[\vzero^T, \vy^T]^T}$. This is a trivial solution, corresponding to $K_p = 0$ and $K_q = \norm{\vy}_0$.
		\FOR{$K_p =1,2,. . .,N$} 
			\FOR{$\l=0,1,. . .,\big(N-1-\tau(S_{\mat{\Psi}}(K_p)-1)\big)$}
				\STATE Use a polynomial-complexity algorithm to estimate a $K_p$-sparse signal $\vx_1$ from a set of consecutive measurements $[y_\l, \ldots, y_{\l + S_{\mat{\Psi}}(K_p) - 1}]$.
	      			\STATE Compute the estimated contribution from the first basis as $\widehat{\vy}= \mat{\Psi} \vx_1$.
				\STATE Compute $\vx_2=\mat{\Phi}^{-1}(\vy-{\widehat{\vy}})$ and let $K_q = \norm{\vx_2}_0$.
	      			\IF{$(K_p, K_q)$ satisfy the condition \eref{other_bases}}
						\STATE Obtain the sparse signal as $\vx = [\vx_1^T, \vx_2^T]^T$.
						\STATE $\mathcal{S} \Leftarrow \mathcal{S} \cup \set{\vx}$.
				\ENDIF
		
			\ENDFOR
		\ENDFOR
	\end{algorithmic}
\end{algorithm}

\subsection{Random Gaussian Basis}
\label{appendix:Gaussian}

Let $\mat{\Psi}_{n, S} \in \R^{S \times N}$ denote a submatrix constructed from $S$ consecutive rows of the random Gaussian matrix $\mat{\Psi}$. The subscript $n$ in $\mat{\Psi}_{n, S}$ indicates that these rows are taken at indices $\set{n, n+1, \ldots, n+S-1}$, where the indices are viewed through the modulo (by $N$) operator. With high probabilities, the normalized matrix $\mat{\Psi}(n, S)/\sqrt{S}$ satisfies the restricted isometry property (RIP) \cite{CandesT:05} in compressed sensing \cite{CandesT:06, Donoho:06, CandesW:08}, allowing one to reconstruct a sparse vector $\vc$ by solving the following convex optimization problem:
\begin{equation}\label{eq:cs_reconstruction}
\text{arg}\,\min_{\widetilde{\vc}} \| \widetilde{\vc} \|_1 \hspace{5mm}
\mbox{s.t.} \hspace{5mm} \mat{\Psi}_{n, S} \,\widetilde{\vc} = \mat{\Psi}_{n, S} \, \vc. 
\end{equation}
More precisely, it was shown in \cite{BaraniukDDW:08} that the following holds: There exist two positive constants $c_1, c_2$ such that, with probability $\ge 1 - 2e^{-c_2 S}$, the matrix $\mat{\Psi}_{n, S}$ will satisfy the RIP with suitable parameters that are sufficient to guarantee the success of the optimization problem \eref{cs_reconstruction} in recovering any $K$-sparse vector $\vc$, for all $K$ satisfying the condition $c_1 K \log(N/K) \le S$.

To apply the result of Proposition~\ref{prop:other_bases}, we need to show that the random matrix $\mat{\Psi}$ will satisfy the following property with high probabilities: One can efficiently reconstruct any $K$-sparse vector $\vc$ from any $S_{\mat{\Psi}}(K)$ consecutive entries of the observation $\vz = \mat{\Psi} \vc$, where $S_{\mat{\Psi}}(K)$ is the function defined in \eref{S_Gaussian}. To that end, it is sufficient to show that, with high probabilities, the submatrices $\mat{\Psi}_{n, S_{\mat{\Psi}}(K)}$ for all $n$ and all $K$ will simultaneously satisfy the RIP condition. We note that any given submatrix $\mat{\Psi}_{n, S_{\mat{\Psi}}(K)}$ will fail to satisfy the required RIP condition with probability $\le 2e^{-c_2 S_{\mat{\Psi}}(K)} <= 2e^{-c_2 p(N)}$. Since there is a total of $N$ possible starting indices (\emph{i.e.}, $0 \le n < N$) and up to $N$ different values of $K$, we can conclude, by applying the union bound, that the matrix $\mat{\Psi}$ will satisfy the desired property with probability at least $1 - 2 N^2 e^{-c_2 p(N)}$. By choosing, for example, $p(N) = (3/c_2) \log(N)$, the previous probability bound becomes $1 - 2 /N$, which can be made arbitrarily close to one for sufficiently large $N$.

\bibliographystyle{IEEEtran}
\bibliography{./biblio/refs_bib}

\begin{thebibliography}{10}
\providecommand{\url}[1]{#1}
\csname url@samestyle\endcsname
\providecommand{\newblock}{\relax}
\providecommand{\bibinfo}[2]{#2}
\providecommand{\BIBentrySTDinterwordspacing}{\spaceskip=0pt\relax}
\providecommand{\BIBentryALTinterwordstretchfactor}{4}
\providecommand{\BIBentryALTinterwordspacing}{\spaceskip=\fontdimen2\font plus
\BIBentryALTinterwordstretchfactor\fontdimen3\font minus
  \fontdimen4\font\relax}
\providecommand{\BIBforeignlanguage}[2]{{%
\expandafter\ifx\csname l@#1\endcsname\relax
\typeout{** WARNING: IEEEtran.bst: No hyphenation pattern has been}%
\typeout{** loaded for the language `#1'. Using the pattern for}%
\typeout{** the default language instead.}%
\else
\language=\csname l@#1\endcsname
\fi
#2}}
\providecommand{\BIBdecl}{\relax}
\BIBdecl

\bibitem{ChenDS:98}
S.~Chen, D.~Donoho, and M.~Saunders, ``Atomic decomposition by basis pursuit,''
  \emph{SIAM J. Scientific Computing}, vol.~20, no.~1, pp. 33--61, Apr. 1998.

\bibitem{DonohoH:01}
D.~Donoho and X.~Huo, ``Uncertainty principle and ideal atomic decomposition,''
  \emph{IEEE Trans. Inf. Theory}, vol.~47, pp. 2845--2862, Nov. 2001.

\bibitem{EladB:02}
M.~Elad and M.~Bruckstein, ``A generalized uncertainty principle and sparse
  representation in pairs of bases,'' \emph{IEEE Trans. Inf. Theory}, vol.~48,
  no.~9, pp. 2558--2567, Sep. 2002.

\bibitem{FeuerN:03}
A.~Feuer and A.~Nemirovsky, ``On sparse representation in pairs of bases,''
  \emph{IEEE Trans. Inf. Theory}, vol.~49, no.~6, pp. 1579--1581, Jun. 2003.

\bibitem{DavisMA:97}
S.~Davis, G.~Mallat and M.~Avallaneda, ``Greedy adaptive approximations,''
  \emph{J. Constr. Approx}, vol.~13, pp. 57--98, 1997.

\bibitem{Natarajan:95}
B.~Natarajan, ``Sparse approximate solutions to linear systems,'' \emph{SIAM J.
  Comput.}, vol.~24, pp. 227--234, 1995.

\bibitem{StoicaM:05}
P.~Stoica and R.~Moses, \emph{Spectral Analysis of Signals}.\hskip 1em plus
  0.5em minus 0.4em\relax Englewood Cliffs, NJ: Prentice-Hall, 2005.

\bibitem{RicaudT:13}
B.~Ricaud and B.~Torr\'esani, ``Refined support and entropic uncertainty
  inequalities,'' \emph{IEEE Trans. Inf. Theory}, vol.~59, no.~7, pp.
  4272--4279, July 2013.

\bibitem{RicaudT:13b}
------, ``A survey of uncertainty principles and some signal processing
  applications,'' \emph{Advances on Computational Mathematics}, October 2013.

\bibitem{Elad:10}
M.~Elad, \emph{Sparse and Redundant Representations}.\hskip 1em plus 0.5em
  minus 0.4em\relax Springer, 2010.

\bibitem{Prony:1795}
G.~C. F. M.~R. Prony, ``Essai exp\'erimental et analytique sur les lois de la
  dilabilit\'e des fluides \'elastiques et sur celles de la force expansive de
  la vapeur de l' eau et de la vapeur de l' alkool, \`a diff\'erentes
  temp\'eratures,'' \emph{J. de l' \'Ecole Polytechnique}, vol.~1, pp. 24--76,
  1795.

\bibitem{Berlekamp:68}
E.~R. Berlekamp, \emph{Algebraic Coding Theory}.\hskip 1em plus 0.5em minus
  0.4em\relax NY: McGraw Hill, 1968.

\bibitem{Massey:69}
J.~L. Massey, ``Shift-register synthesis and {BCH} decoding,'' \emph{IEEE
  Trans. Inf. Theory}, vol.~15, no.~1, pp. 122--127, Jan. 1969.

\bibitem{MilanfarVKW:95}
P.~Milanfar, G.~C. Verghese, W.~Karl, and A.~S. Willsky, ``Reconstructing
  polygons from moments with connections to array processing,'' \emph{IEEE
  Trans. on Signal Processing}, vol. 43(2), pp. 432--443, February 1995.

\bibitem{GustafssonHMP:00}
B.~Gustafsson, C.~He, P.~Milanfar, and M.~Putinar, ``Reconstructing planar
  domains from their moments,'' \emph{Inverse Prob.}, vol.~16, no.~54, pp.
  1053--1070, Aug. 2000.

\bibitem{EladMG:04}
M.~Elad, P.~Milanfar, and G.~H. Golub, ``Shapes from moments --- {A}n
  estimation theory perspective,'' \emph{IEEE Trans. Signal Process.}, vol.~52,
  no.~7, pp. 1814--1829, Jul. 2004.

\bibitem{DragottiVB:07}
P.~Dragotti, M.~Vetterli, and T.~Blu, ``Sampling moments and reconstructing
  signals of finite rate of innovation: {S}hannon meets {S}trang-{F}ix,''
  \emph{IEEE Trans. Signal Process.}, vol.~55, no.~5, pp. 1741--1757, May 2007.

\bibitem{VetterliMB:02}
M.~Vetterli, P.~Marziliano, and T.~Blu, ``Sampling signals with finite rate of
  innovation,'' \emph{IEEE Trans. Signal Process.}, vol.~50, no.~6, pp.
  1417--1428, Jun. 2002.

\bibitem{HormatiRLM:10}
A.~Hormati, O.~Roy, Y.~M. Lu, and M.~Vetterli, ``Distributed sampling of
  correlated signals linked by sparse filtering: {T}heory and applications,''
  \emph{IEEE Trans. Signal Process.}, vol.~58, no.~3, pp. 1095--1109, Mar.
  2010.

\bibitem{CandesT:06}
E.~J. Cand\`{e}s and T.~Tao, ``Near optimal signal recovery from random
  projections: Universal encoding strategies?'' \emph{IEEE Trans. Inf. Theory},
  vol.~52, no.~12, pp. 5406--5425, Dec. 2006.

\bibitem{Donoho:06}
D.~Donoho, ``Compressed sensing,'' \emph{IEEE Trans. Inf. Theory}, vol.~52,
  no.~4, pp. 1289--1306, Apr. 2006.

\bibitem{BaraniukDDW:08}
R.~Baraniuk, M.~Davenport, R.~{DeVore}, and M.~Wakin, ``A simple proof of the
  restricted isometry property for random matrices,'' \emph{Constructive
  Approximation}, vol.~28, no.~3, pp. 253--263, 2008.

\bibitem{CandesT:05}
E.~J. Cand\`{e}s and T.~Tao, ``Decoding by linear programming,'' \emph{IEEE
  Trans. Inf. Theory}, vol.~51, pp. 4203--4215, Dec. 2005.

\bibitem{CandesW:08}
E.~J. Cand\`{e}s and M.~B. Wakin, ``An introduction to compressive sampling,''
  \emph{IEEE Signal Process. Mag.}, vol.~21, Mar. 2008.

\end{thebibliography}

\end{document}